\documentclass[12p]{elsarticle}

\renewcommand{\footnotesize}{\scriptsize}
\usepackage{textcomp}
\usepackage{listings} 
\usepackage{graphicx}
\usepackage{caption}
\usepackage{subcaption}

\usepackage{csquotes}

\usepackage{tikz}
\usetikzlibrary{tikzmark}
\usetikzlibrary{fit} 
\usetikzlibrary{positioning}
\usetikzlibrary{arrows}
\usetikzlibrary{shapes.multipart}

\usepackage[hidelinks,bookmarks=false]{hyperref}
\usepackage[numbered]{bookmark}
\usepackage{color,soul}
\usepackage{booktabs}
\usepackage{multirow}
\usepackage{url}
\usepackage{tcolorbox}
\usepackage{amsmath,amssymb,amsfonts}
\usepackage{algorithmic}
\usepackage{xcolor}
\usepackage{url}
\usepackage{tabularx} 
\usepackage{array}
\usepackage{lineno,hyperref}
\usepackage{longtable}
\modulolinenumbers[5]
\journal{Journal of \LaTeX\ Templates}

\usepackage{pgfplots}
\pgfplotsset{compat=1.14}

\usepackage{adjustbox}
\usepackage{float}
\usepackage{rotating}
\usepackage{tablefootnote}
\usepackage{nth}
\DeclareCaptionType{TextBox}
\newcolumntype{L}{>{\centering\arraybackslash}m{3cm}}

\tcbset{tab1/.style={fonttitle=\bfseries\large,fontupper=\normalsize\sffamily,
colback=yellow!6!white,colframe=red!75!black,colbacktitle=black!75!black,
coltitle=white,center title,freelance,frame code={
\foreach \n in {north east,north west,south east,south west}
{\path [fill=red!75!black] (interior.\n) circle (3mm); };},}}

\tcbset{tab2/.style={enhanced,fonttitle=\bfseries,fontupper=\tiny\sffamily,
colback=yellow!6!white,colframe=red!50!black,colbacktitle=black!75!black,
coltitle=white,center title}}
\usepackage[T1]{fontenc}
\usepackage{array, booktabs, makecell}
\usepackage[utf8]{inputenc}
\usepackage{dirtytalk}
\usepackage{tcolorbox}

\usepackage[british]{babel}
\usepackage{enumitem}

\newlist{SubItemList}{itemize}{1}
\setlist[SubItemList]{label={$-$}}

\let\OldItem\item
\newcommand{\SubItemStart}[1]{%
    \let\item\SubItemEnd
    \begin{SubItemList}[resume]%
        \OldItem #1%
}
\newcommand{\SubItemMiddle}[1]{%
    \OldItem #1%
}
\newcommand{\SubItemEnd}[1]{%
    \end{SubItemList}%
    \let\item\OldItem
    \item #1%
}
\newcommand*{\SubItem}[1]{%
    \let\SubItem\SubItemMiddle%
    \SubItemStart{#1}%
}

\usepackage{framed}
\usepackage{mdframed}
\usepackage{pstricks}
\usepackage{xspace}
\usepackage{pgf-pie}
\usepackage{comment}
\usepackage{colortbl}
\tcbuselibrary{skins}
\tcbuselibrary{listings}
\usepackage[normalem]{ulem}
\usetikzlibrary{patterns,}
\usepgfplotslibrary{colorbrewer}
\usepackage{verbatim}
\usepackage{tikz}
\usetikzlibrary{tikzmark}
\usetikzlibrary{fit} 
\usetikzlibrary{positioning}
\usetikzlibrary{arrows}
\usetikzlibrary{shapes.multipart}
\usepackage{lscape}
\usepackage{placeins}
\usepackage{afterpage}
\newlength\mylength
\usepackage{fontawesome}
\usepackage{xspace}
\newcommand{\ie}{\textit{i.e., \xspace}}
\newcommand{\eg}{\textit{e.g., \xspace}}
\newcommand{\etal}{\textit{et al. \xspace}}
\newcolumntype{L}{>{\arraybackslash}m{16cm}}
\usepackage[flushleft]{threeparttable}
\usepackage{mdframed}

\newtcolorbox{boxK}{
    sharpish corners, 
    boxrule = 0pt,
    toprule = 4.5pt, 
    enhanced,
    fuzzy shadow = {0pt}{-2pt}{-0.5pt}{0.5pt}{black!35} 
}
\begin{document}

\begin{frontmatter}

\title{An Empirical Study on the Impact of Code Duplication-aware Refactoring Practices on Quality Metrics}

\author[STEVENS]{Eman Abdullah AlOmar\corref{mycorrespondingauthor}}
\cortext[mycorrespondingauthor]{Corresponding author}
\ead{ealomar@stevens.edu}


\address[STEVENS]{Stevens Institute of Technology, Hoboken, NJ, USA}

\begin{abstract}
\textbf{Context:} Code refactoring is widely recognized as an essential software engineering practice that improves the
understandability and maintainability of source code. Several studies attempted to detect refactoring activities through mining software repositories, allowing one to collect, analyze, and get actionable data-driven insights about refactoring practices within software projects. 

\noindent\textbf{Objective:} Our goal is to identify, among the various quality models presented in the literature, the ones that align with the developer's vision of eliminating duplicates of code, when they explicitly mention that they refactor the code to improve them. 

\noindent\textbf{Method:} We extract a corpus of 332 refactoring commits applied and documented by developers during their daily changes from 128 open-source Java projects. In particular, we extract 32 structural metrics 
from which we identify code duplicate removal commits with their corresponding refactoring operations, as perceived by software engineers. Thereafter, we empirically analyze the impact of these refactoring operations on a set of common state-of-the-art design quality metrics. 

\noindent\textbf{Results:} The statistical analysis of the results obtained shows that (i) some state-of-the-art metrics are capable of capturing the developer's intention of removing code duplication; and
(ii) some metrics are being more emphasized than others. We confirm that various structural metrics can effectively represent code duplication, leading to different impacts on software quality. Some metrics contribute to improvements, while others may lead to degradation. 

\noindent\textbf{Conclusion:} Most of the mapped metrics associated with the main quality attributes successfully capture developers' intentions for removing code duplicates, as is evident from the commit messages. However, certain metrics do not fully capture these intentions.

\end{abstract}

\end{frontmatter}


\section{Introduction}
\label{Section:Introduction}

Duplicating a code fragment involves the process of copying and pasting it, with or without minor modifications, into another section of the codebase. Although this may appear to be an intuitive way to reuse code, the presence of duplicate code introduces a set of challenges in the maintenance and evolution of software systems~\citep{roy2009comparison}. Recent studies emphasize that duplicate code has become a significant issue that affects both developers and researchers. For example, fixing a bug in a duplicate code fragment may require applying the same fix to all instances of that code \citep{thongtanunam2019will}. This can result in a slowdown of maintenance efforts and potentially lead to the widespread propagation of bugs throughout the codebase. Consequently, in response to these challenges, the elimination of duplicate code through refactoring has become a natural and necessary course of action \citep{fanta1999removing}.

Refactoring is the art of remodeling software design without altering its functionality \citep{Fowler:1999:RID:311424}. It is a critical software maintenance activity that developers perform for a variety of purposes: improve program comprehension, eliminate duplicate code, reduce complexity, manage technical debt, and remove code smells~\citep{silva2016we,murphy2008breaking}. Refactoring duplicate code involves the process of taking a code fragment and moving it to create a new method while replacing all instances of that fragment with a call to this newly created method.

Despite the increasing emphasis on recommending refactorings through the optimization of structural metrics and the removal of code smells, there is limited evidence on whether developers follow these recommendations when refactoring their code. A study by Pantiuchina \etal \citep{pantiuchina2018improving} has shown that there is a misperception between state-of-the-art structural metrics, widely used as indicators for refactoring, and what developers actually consider to be an improvement in their source code. Furthermore, AlOmar \etal \citep{alomar2019impact} reveals that most metrics that are mapped to the main quality
attributes, \ie cohesion, coupling, and complexity, capture the developer's intentions of quality improvement reported in the
commit messages. In contrast, there is also a case in which
the metrics do not capture quality improvement as perceived
by developers. \textcolor{black}{This paper builds upon our previously published paper \citep{alomar2019impact}. Although earlier research primarily focused on internal quality attributes, placing particular emphasis on whether developer intentions regarding cohesion, coupling, complexity, inheritance, polymorphism, encapsulation, abstraction, and size aligned with their vision of quality optimization, this paper presents a more comprehensive qualitative and quantitative approach with a focus on code duplication. Code duplication is a critical quality issue, especially with recent emphasis on developing tools for this purpose. Furthermore, a study involving GitHub contributors has shown that they are seriously concerned about code duplication \citep{silva2016we}. Specifically, we extend the study by}:
 \begin{itemize}
     \item \textcolor{black}{Examining which metrics are most impacted by refactorings, aiming to identify those that closely capture the developer's intention regarding code duplication rather than internal quality attributes.}
    \item \textcolor{black}{Providing numerous qualitative examples that offer deeper insights into the underlying reasons for instances of alignment and disparity between quality metrics and developers' perception of the removal of duplicate code.}
   \item \textcolor{black}{Offering lessons and insights derived from our experiments to developers, tool builders, and the research community, contributing to the advancement of both the state-of-the-art and state-of-the-practice in code duplication-aware refactoring practices.}
   \item \textcolor{black}{Utilizing an entirely different dataset than the previous work.}
\item \textcolor{black}{Leveraging 32 structural metrics, which encompasses 22 distinct metrics that were not employed in the previous paper.}
     \end{itemize}






We start with reviewing studies from the literature that propose quality attributes of software and their corresponding measurement in the source code, in terms of metrics. Software quality attributes are typically characterized by high-level definitions whose interpretations allow multiple ways to calculate them in the source code. Thus, there is little consensus on what would be the optimal match between quality attributes and code-level design metrics. For instance, as shown later in Section \ref{Section:Background}, the notion of complexity was the subject of many studies that proposed several metrics to calculate it. Therefore, we investigate which code-level metrics are more representative of the removal of high-level code duplicates, when their optimization is explicitly stated by the developer when applying refactorings. Furthermore, we investigate the performed refactoring operations, for each explicitly mentioned removal of code duplication.



Practically, we have classified 322 commits, as duplicated code removal commits, by manually analyzing their messages and identifying an explicit statement of removing duplicated code, along with detecting their refactoring activities. We use the SmartSHARK dataset \citep{trautsch2021msr}, and refine it by untangling each commit to select only refactored code elements. Afterward, we calculate the values of its corresponding structural metrics, in the files, before and after their refactorings. And finally, we empirically compare the variation of these values, to distinguish the metrics that are significantly impacted by the refactorings, and so they better reflect the developer's intention of removing code duplication. To the best of our knowledge, no previous study has investigated the relationship between the intention of code duplicate removal and their corresponding structural metrics, from the developer's perception. Our key findings show that not all state-of-the-art structural metrics equally represent code duplication. 
 In summary, this extended paper makes the following key contributions:

 \begin{itemize}
       
     \item In our empirical investigation into the removal of duplicate code, we examine which metrics are most affected by refactorings, aiming to identify those that closely capture the developer's intention. 
\item We provide numerous qualitative examples that offer deeper insights into the underlying reasons for instances of alignment and disparity between quality metrics and developers' perception of the removal of duplicate code.
\item We offer lessons and insights derived from our experiments to developers, tool builders, and the research community, aiming to contribute to the advancement of both the state-of-the-art and state-of-the-practice in code duplication-aware refactoring practices.
\item We provide our  experiment package to further replicate and extend our study. The package contains the raw data, analyzed data, statistical test results, and custom scripts used in our research\footnote{\textcolor{black}{\url{https://smilevo.github.io/self-affirmed-refactoring/}}}. 
 \end{itemize}
The remainder of this paper is organized as follows. Section \ref{Section:Background} reviews existing studies related to the impact of refactoring on quality. Section \ref{Section:methodology} outlines our empirical setup in terms of data extraction, analysis, and research question. Section \ref{Section:Result} discusses our findings, while the \textcolor{black}{lessons learned and } implications of the research are discussed in Sections \ref{Section:lesson} and \ref{Section:Implication}, respectively. Section \ref{Section:Threats} captures any threats to the validity of our work, before concluding with Section \ref{Section:Conclusion}.
\section{Related Work}
\label{Section:Background}

\begin{table*}
  \centering
	 \caption{\textcolor{black}{A summary of the literature on the impact of refactoring activities on software quality attributes.}}
	 \label{Table:Quality Metrics in Related Work}
\begin{adjustbox}{width=1.6\textwidth,center}
\rowcolors{2}{gray!25}{white}
\begin{tabular}{llllll}\hline
\toprule
 \bfseries No. & \bfseries Study & \bfseries Year & \bfseries Quality Metric & \bfseries Internal QA & \bfseries External QA  \\
\midrule

1 & Sahraoui \etal \cite{sahraoui2000can} & 2000  & CLD / NOC / NMO / NMI    & Inheritance / Coupling & Fault-proneness / Maintainability   \\ 
& & &  NMA / SIX / CBO / DAC & & \\
& & &  IH-ICP / OCAIC / DMMEC / OMMEC &  \\ \hline
2 & Stroulia \& Kapoor \cite{stroulia2001metrics} & 2001 & LOC / LCOM / CC & Size / Coupling & Design extensibility \\ \hline
3 & Kataoka \etal \cite{kataoka2002quantitative} & 2002 & Coupling measures &  Coupling & Maintainability   \\ \hline
4 & Demeyer \cite{demeyer2002maintainability} & 2002& N/A & Polymorphism  & Performance  \\ \hline
5 & Tahvildari \etal \cite{tahvildari2003quality} & 2003 & LOC / CC / CMT / Halstead's efforts & Complexity  & Performance / Maintainability   \\ \hline
6 & Leitch \& Stroulia \cite{leitch2003assessing}& 2003 & SLOC / No. of Procedure & Size & Maintainability  \\ \hline
7 & Bois \& Mens \cite{du2003describing} & 2003  & NOM / CC / NOC / CBO & Inheritance / Cohesion / Coupling / Size / Complexity & N/A  \\ 
& & &  RFC / LCOM & & \\ \hline
8 & Tahvildari \& Kontogiannis \cite{tahvildari2003metric} & 2004  & LCOM / WMC / RFC / NOM   & Inheritance / Cohesion / Coupling / Complexity & Maintainability  \\ 
& & &  CDE / DAC / TCC & & \\ \hline
9 & Bois \etal \cite{du2004refactoring} & 2004  & N/A & Cohesion / Coupling & Maintainability   \\ \hline
10 & Bois \etal \cite{du2005does} & 2005  & N/A & N/A &   Understandability   \\  \hline
11 & Geppert \etal \cite{geppert2005refactoring} & 2005  &  N/A & N/A & Changeability  \\ \hline
12 & Ratzinger \etal \cite{ratzinger2005improving} & 2005  & N/A &  Coupling & Evolvability \\ 
& & & Analyzing code histories & \\ \hline
13 & Moser \etal \cite{moser2006does} & 2006  & CK / MCC / LOC   & Inheritance / Cohesion / Coupling / Complexity & Reusability \\ \hline
14 & Wilking \etal \cite{wilking2007empirical} & 2007  & CC / LOC  & Complexity & Maintainability / Modifiability   \\ \hline
15 & Stroggylos \& Spinells \cite{stroggylos2007refactoring} & 2007  & CK / Ca / NPM & Inheritance / Cohesion / Coupling / Complexity & N/A  \\ \hline
16 & Moser \etal \cite{moser2007case} & 2008 & CK / LOC / Effort (hour) & Cohesion / Coupling / Complexity & Productivity  \\ \hline
17 & Shrivastava \& Shrivastava \cite{shrivastava2008impact} & 2008 & NOA / NOC / NOM / CC & Inheritance / Complexity / Size & N/A\\ 
& & & TLOC / DIT & \\ \hline 
18 & Higo \etal \cite{higo2008refactoring} & 2008 & CK & Inheritance / Cohesion / Coupling / Complexity & N/A\\ \hline
19 & Reddy \& Rao  \cite{Reddy2009quantitative} & 2009 & DOCMA (CR) \ DOCMA (AR) & Complexity & N/A \\ \hline
20 & Alshayeb \cite{alshayeb2009empirical} & 2009 &   CK / LOC / FANOUT  & Inheritance / Cohesion / Coupling / Size & Adaptability / Maintainability / Testability / Reusability  \\ 
& & & & &  Understandability  \\ \hline
21 & Alshayeb \cite{alshayeb2009refactoring} & 2009 & LCOM1 / LCOM2 / LCOM3 / LCOM4 / LCOM5 & Cohesion & N/A \\ \hline
22 & Usha \etal \cite{usha2009quantitative} & 2009 & LCOM / CBO / WMC / RFC / CC & Cohesion / Coupling / Complexity & Modifiability / Modularity \\ 
& & &  CF / TCC / MHF / AHF & & \\ \hline 
23 & Hegedus \etal \cite{hegedHus2010effect} & 2010  & CK  & Coupling / Complexity / Size & Maintainability / Testability / Error Proneness / Changeability  \\
& & & & &  Stability / Analizability \\ \hline
24 & Shatnawi \& Li \cite{shatnawi2011empirical} & 2011 & CK / QMOOD &  Inheritance / Cohesion / Coupling / Polymorphism / Size & Reusability / Flexibility / Extendibility / Effectiveness     \\ 
& & & & Encapsulation / Composition / Abstraction / Messaging &    \\ \hline
25 & Fontana \& Spinelli \cite{fontana2011impact} & 2011   & DAC / LCOM / NOM / RFC & Cohesion / Coupling / Complexity & N/A\\ 
& & &  TCC / WMC & &  \\ \hline 
26 & Alshayeb \cite{alshayeb2011impact} & 2011  &  CK / LOC / FANOUT  & Inheritance / Cohesion / Coupling / Size & Adaptability / Maintainability / Testability / Reusability  \\ 
& & & & &  Understandability \\ \hline
27 & Lerthathairat \& Prompoon \cite{lerthathairat2011approach} & 2011 & NLOC / NILI / CC / ILCC / NOP & Cohesion / Encapsulation   & N/A \\
& & & NOM / NFD / LCOM / LCOM-HS &  & \\ \hline
28 & {\'O} Cinn{\'e}ide \etal \cite{o2012experimental} & 2012 & LSCC / TCC / CC / SCOM / LCOM5 & Cohesion & N/A \\ \hline 
29 & Ibrahim \etal \cite{ibrahim2012identification} & 2012 & LCCI / LCCD / LCC / TCC  & Cohesion & N/A \\
& & & CC / Coh / LCOM3 &  \\ \hline 
30 & Singh \& Kahlon \cite{singh2011effectiveness} & 2011 &  CK / LCOM4 / PuF / EncF / NOD & Inheritance / Cohesion / Coupling / Information hiding &  \\
& & & & Polymorphism / Encapsulation / Abstraction & \\\hline
31 & Singh \& Kahlon \cite{singh2012effectiveness} & 2012 & CK / LCOM4 / PuF / EncF / NOD & Inheritance / Cohesion / Coupling / Information hiding &  \\
& & & & Polymorphism / Encapsulation / Abstraction & \\\hline
32 & Murgia \etal \cite{murgia2012refactoring} & 2012 & FANIN / FANOUT & Coupling & N/A \\ \hline 
33 & Kannangara \& Wijayanake \cite{kannangara2013impact} & 2013 &  N/A  & N/A & Analysability / Changeability / Time Behaviour / Resource Utilization  \\ \hline 
34 & Veerappa \& Rachel \cite{veerappa2013empirical} & 2013 & RFC / DCC / CBO / MPC & Coupling & N/A \\ \hline
35 & Napoli \etal \cite{napoli2013using} & 2013 & LCOM / CBO & Cohesion / Coupling  & Modularity \\ \hline 
36 & Bavota \etal \cite{bavota2013empirical} & 2013  & ICP / IC-CD / CCBC & Coupling & N/A \\
& &  & & & \\ \hline
37 & Kumari \& Saha \cite{kumari2014effect} & 2014 & DIT / CBO / RFC / WMC  & Inheritance / Cohesion / Coupling / Complexity & Maintainability / Reusability / Testability / Understandability \\ 
& & & LCOM / NOM / LOC &  &  Fault proneness / Completeness / Stability / Adaptability \\ \hline 
38 & Szoke \etal \cite{szoke2014bulk} & 2014 & CC / U / NOA / NII / NAni & Size / Complexity & N/A \\
& &  & LOC / NUMPAR / NMni / NA & &  \\ \hline
39 & Chaparro \etal \cite{chaparro2014impact} & 2014 &  RFC / CBO / DAC / MPC & Inheritance / Cohesion / Coupling / Size / Complexity & N/A\\ 
& &  & LOC / NOM / CC / LCOM2 & &  \\ 
& & & LCOM5 / NOC / DIT & & \\ \hline
40 & Bavota \etal \cite{bavota2015experimental} & 2015  &  CK / LOC / NOA / NOO  &  Inheritance / Cohesion / Coupling / Size / Complexity & N/A \\
& &  & C3 / CCBC & & \\ \hline
41 & Kannangara \& Wijayanake \cite{kannangara2015empirical} & 2015 &  CC / DIT / CBO / LOC & Maintainability index / Complexity / Coupling / Inheritance & Analysability / Changeability / Time Behaviour / Resource Utilization  \\ \hline 
42 & Gatrell \&  Counsel \cite{gatrell2015effect} & 2015 & N/A & N/A & change \& fault-proneness \\ \hline 
43 & Cedrim at al. \cite{cedrim2016does} & 2016 & LOC / CBO / NOM / CC & Cohesion / Coupling / Complexity & N/A  \\
& &  & FANOUT / FANIN & &  \\ 

\bottomrule
\end{tabular}
\end{adjustbox}
\end{table*}
\begin{table*}
  \centering
	 \caption{\textcolor{black}{Continued from previous page.}}
	 \label{Table:Quality Metrics in Related Work-v2}
\begin{adjustbox}{width=1.6\textwidth,center}
\rowcolors{2}{gray!25}{white}
\begin{tabular}{llllll}\hline
\toprule
 \bfseries No. & \bfseries Study & \bfseries Year & \bfseries Quality Metric & \bfseries Internal QA & \bfseries External QA  \\
\midrule

44 & Malhotra \& Chug \cite{malhotra2016empirical} & 2016 & CK & Cohesion / Coupling / Complexity / Inheritance & Understandability / Modifiability / Extensibility / Reusability   \\ 
& & & & & Level of Abstraction \\ \hline 
45 & Mkaouer \etal \cite{mkaouer2016use} & 2016 & QMOOD & N/A & Reusability / Flexibility / Understandability / Functionality\\ 
& & & && Extendibility / Effectiveness \\ \hline 
46 & Kaur \& Singh \cite{kaur2017improving} & 2017 & WMC / NOI / RFC / TCLOC  & Coupling / Complexity / Size  & Maintainability \\ 
& & & TLLOC / TNOS /  CI  & & \\ \hline 
47 & Chavez \etal \cite{chavez2017does} & 2017  & CBO / WMC / DIT / NOC & Inheritance / Cohesion / Coupling / Size / Complexity & N/A  \\ 
&& &   LOC / LCOM2 / LCOM3 / WOC & &  \\
&& &  TCC / FANIN / FANOUT / CINT & &  \\
&& &  CDISP / CC / Evg / NPATH   & & \\
&& &  MaxNest / IFANIN / OR / CLOC & &  \\
&& & STMTC / CDL / NIV / NIM / NOPA & &  \\ \hline 
48 & Szoke \etal \cite{szHoke2017empirical} & 2017 & CK & N/A & Maintainability \\ \hline
49 & Bashir \etal \cite{bashir2017methodology} & 2017 & QMOOD & N/A & Modifiability / Analyzability / Understandability / Maintainability \\ \hline 
50 & Mumtaz \etal \cite{mumtaz2018empirical} & 2018 &  CK / CCC / CDP / CCDA / COA & N/A & Security \\ 
& & &  CMW / CMAI / CAAI / CAIW & &  \\ \hline 
51 & Pantiuchina \etal \cite{pantiuchina2018improving} & 2018 &  LCOM / CBO / WMC / RFC  & Cohesion / Coupling / Complexity & Readability  \\
& &   &  C3 / B\&W / SRead & & \\ \hline 
52 & Alizadeh \& Kessentini \cite{alizadeh2018reducing} & 2018 & QMOOD & N/A & Reusability / Flexibility / Understandability / Functionality \\ 
& & & && Extendibility / Effectiveness   \\ \hline 
53 & Alizadeh \etal \cite{alizadeh2019refbot} & 2019 & QMOOD & N/A & Reusability / Flexibility / Understandability / Functionality \\ 
& & & && Extendibility / Effectiveness   \\ \hline 
54 & Techapalokul \& Tilevich \cite{techapalokul2019code} & 2019 & LOC / Complex Script Dens / No. Literals & N/A & N/A  \\ 
& & & Long Script Dens. / Procedure Dens. / No. Global Var & & \\
& & & No. Create Clone Of. & & \\ \hline 
55 & Counsell \etal \cite{counsell2019relationship} & 2019 & CBO & Coupling & N/A  \\ \hline 
56 & Fakhoury \etal \cite{fakhoury2019improving} & 2019 & Buse \& Weimer / Dorn / Scalabrino / Posnett & Cohesion / Coupling / Size / Complexity  & Readability \\ 
& & & LCOM 5 / WMC / RFC / MLOC / FLOC  & &  \\ 
& & & Halstead Difficulty / Halstead Effort / Maintainability index / MCC & &  \\ 
& & & Nesting level / Doc LOC / Comment Density / API Doc & & \\ 
& & & Public Undoc API / Public Doc API / \# Parantheses / & &  \\ 
& & & Number of Incoming Invocations & &  \\ \hline 
57 & AlOmar \etal \cite{alomar2019impact} & 2019 &  CK / FANIN / FANOUT / CC / NIV / NIM & Inheritance / Cohesion / Coupling / Complexity  & N/A \\ 
& &   & Evg / NPath / MaxNest / IFANIN & Size / Polymorphism / Encapsulation / Abstraction &   \\ 
& & &  LOC / CLOC / CDL / STMTC & &  \\ \hline 
58 & Rebai \etal \cite{rebai2019interactive} & 2019 & QMOOD & N/A & Reusability / Flexibility / Understandability / Functionality \\ 
& & & && Extendibility / Effectiveness   \\ \hline 
59 & Alizadeh \etal \cite{alizadeh2019less} & 2019 & QMOOD & N/A & Reusability / Flexibility / Understandability / Functionality \\ 
& & & && Extendibility / Effectiveness   \\ \hline 
60 & Alizadeh \etal \cite{alizadeh2018interactive} & 2020 & QMOOD & N/A & Reusability / Flexibility / Understandability / Functionality \\ 
& & & && Extendibility / Effectiveness   \\ \hline 
61 & Rebai \etal \cite{rebai2020enabling} & 2020 & QMOOD & N/A & Reusability / Flexibility / Understandability / Functionality \\ 
& & & && Extendibility / Effectiveness   \\ \hline 
62 & Fernandes \etal \cite{fernandes2020refactoring} & 2020    & CBO / WMC / DIT / NOC & Inheritance / Cohesion / Coupling / Size / Complexity & N/A  \\
&&  & LOC / LCOM2 / LCOM3 / WOC & &  \\
&& &  TCC / FANIN / FANOUT / CINT & &  \\
&& &  CDISP / CC / Evg / NPATH   & & \\
&& &  MaxNest / IFANIN / OR / CLOC & &  \\
&& &  STMTC / CDL / NIV / NIM / NOPA & &  \\ \hline
63 & AlOmar \etal \cite{alomar2020developers} & 2020 & CK /  CC / LOC & Inheritance / Cohesion / Coupling / Complexity / Size  & Reusability  \\ \hline
64 & Bibiano \etal \cite{bibiano2020does} & 2020 & LCOM2 / CBO / MAXNest / CC &  Cohesion / Coupling / Complexity / Size & N/A \\ 
& & & LOC / CLOC  / STMTC / NIV & &  \\ 
&& &  NIM / WMC & &  \\ \hline 
65 & Abid \etal \cite{abid2020does} & 2020 & QMOOD & N/A & Reusability / Flexibility / Understandability / Functionality \\ & & & && Extendibility / Effectiveness / Security  \\ \hline 
66 & Abid \etal \cite{abid2021prioritizing} & 2021 & QMOOD & N/A & Reusability / Flexibility / Understandability / Functionality  \\ 
& & & & & Extendibility / Effectiveness / Security   \\ \hline  
67 & Riansyah \& Mursanto \cite{riansyah2020empirical} & 2020 & CINT / CDISP & Coupling & N/A \\ \hline
68 & Iyad \etal \cite{alazzam2020impact} & 2020 & CK / CC / TLOC / MFA / NBD & Inheritance / Cohesion / Coupling / Complexity / Size & N/A \\
& & & NSC / CE & &  \\ \hline
69 & Hamdi \etal \cite{hamdi2021empirical} & 2021  & CBO / WMC / DIT / RFC & Inheritance / Cohesion / Coupling / Complexity / Size  & N/A\\ 
& &    & LCOM / TCC / LOC / LCC & &  \\ 
& & &  NOSI / VQTY & &  \\ \hline
70 & AlOmar \etal \cite{alomar2022refactoring} & 2021  & CK /  CC / LOC / NPATH & Inheritance / Cohesion / Coupling / Complexity / Size  & Reusability  \\ 
& & &  MaxNest / IFANIN / CDL / CLOC & &  \\
& & &  FANIN / FANOUT / STMTC / NIV & &  \\ \hline 
71 & Sellitto \etal \cite{sellittotoward} & 2021 & CIC / CIC\_syc / ITID / NMI / CR & N/A & Readability \\
& & & NM / TC / NOC / NOC\_norm & & \\ \hline
72 & Ouni \etal \cite{ouni2023impact}    & 2023  & LCOM / CBO / NOSI / TCC / NIV / IFANIN& Coupling / Cohesion / Complexity / Inheritance / Size & N/A\\ 
&  & & RFC / FANIN / WMC / VQYT / NIM & & \\
&  &  & FANOUT / CC / Evg / MaxNest / DIT & & \\
& & &  LOC / BLOC / CLOC / STMTC / NOC & & \\

\bottomrule
\end{tabular}
\end{adjustbox}
\end{table*}

The prevailing consensus in the software refactoring literature acknowledges its overarching aim of improving software quality and correcting poor design and implementation practices \citep{Fowler:1999:RID:311424}. \textcolor{black}{\textcolor{black}{Tables \ref{Table:Quality Metrics in Related Work} and \ref{Table:Quality Metrics in Related Work-v2}} illustrate two decades of work on a long-standing question within the refactoring community: Does refactoring improve code quality? In recent years, numerous research efforts have been made to examine and explore the influence of refactoring on software quality} \citep{moser2007case,wilking2007empirical,alshayeb2009empirical, shatnawi2011empirical,bavota2015experimental, chavez2017does,mkaouer2017robust, cedrim2016does,hegedHus2010effect}. 
Most studies have focused on measuring internal and external quality attributes to determine the quality of a software system being refactored. \textcolor{black}{Due to space constraints, this section provides a comprehensive review of some of these studies and a discussion of the pertinent literature on the impact of refactoring on software quality.}

Stroulia and Kapoor \citep{stroulia2001metrics} explored the effect of size and coupling measures on software quality after the refactoring application. Their findings indicated that size and coupling metrics decreased after refactorings. Fioravanti \etal \citep{fioravanti2001reengineering} analyzed and described metrics, based on duplication analysis, that
contribute to the process of reengineering analysis of
object-oriented. Antoniol \etal \citep{antoniol2002analyzing} studies cloning evolution in the Linux kernel. Their main result revealed that the Linux system does not contain a relevant fraction of code
duplication.  Kataoka \etal \citep{kataoka2002quantitative} focused solely on coupling measures to study the impact of \textit{Extract Method} and \textit{Extract Class} refactoring operations on the maintainability of a C++ software system. Their study revealed a positive effect of refactoring on system maintainability. Demeyer \citep{demeyer2002maintainability} conducted a comparative study to investigate the impact of refactoring on performance, the results demonstrating an improvement in program performance after refactoring. In addition, Sahraoui \etal \citep{sahraoui2000can} used coupling and inheritance measures to automatically identify potential antipatterns and predict scenarios in which refactoring could enhance software maintainability. The authors found that quality metrics can help bridge the gap between design improvement and automation, but in some situations the process cannot be fully automated, as it requires the programmer's validation through manual inspection. 

Tahvildari \etal \citep{tahvildari2003quality} introduced a software transformation framework that connects software quality requirements, such as performance and maintainability, with the transformation of the program to enhance the targeted qualities. Their results showed that utilizing design patterns increases the system's maintainability and performance. In a related study, Tahvildari and Kontogiannis \citep{tahvildari2003metric} applied the same framework to assess four object-oriented measures (cohesion, coupling, complexity, and inheritance) together with software maintainability. Leitch and Stroulia \citep{leitch2003assessing} utilized dependency graph-based techniques to investigate the impact of two refactorings, namely, \textit{Extract Method} and \textit{Move Method}, on software maintenance using two small systems. Their findings demonstrated that refactoring improved quality by reducing the size of the design, increasing the number of procedures, decreasing data dependencies and minimizing regression testing. Bios and Mens \citep{du2003describing} proposed a framework to analyze the impact of three refactorings on five internal quality attributes (\textit{i.e.,} cohesion, coupling, complexity, inheritance, and size). Their results indicated both positive and negative impacts on the selected measures.  Bios \etal \citep{du2004refactoring} provided a set of guidelines to optimize cohesion and coupling measures. Their study showed that the impact of refactoring on these measures ranged from negative to positive. In a subsequent study,  Bios \etal \citep{du2005does} differentiated between the application of Refactor to Understand and the traditional Read to Understand pattern, demonstrating that refactoring plays a role in improving software understandability. Rieger \etal \citep{rieger2004insights} provided insight into system-wide code duplication. The author proposed a way of grouping the duplication information
into useful abstractions and proposed a number of
polymetric views that structure the data and combine it
with the knowledge about the system that the engineer possesses.

Geppert \etal \citep{geppert2005refactoring} investigated the impact of refactoring on changeability by focusing on three factors: customer-reported defect rates, change effort, and scope of changes. Their findings showed a significant decrease in the first two factors. Ratzinger \etal \citep{ratzinger2005improving} analyzed historical data from a large industrial system, focusing on reducing change couplings. By examining identified change couplings and corresponding code smell changes, they determined efficient areas for applying refactoring, concluding that refactoring has the potential to enhance software evolvability, specifically by reducing change coupling. In an agile development environment, Moser \etal \citep{moser2006does} used internal measures (\textit{i.e.,} CK, MCC, LOC) to explore the effect of refactoring on the reusability of the code using a commercial system. Their study indicated that refactoring could enhance the reusability of classes that initially were difficult to reuse. Wilking \etal \citep{wilking2007empirical} empirically studied the effect of refactoring on non-functional aspects, \textit{i.e.,} the maintainability and modifiability of system systems. They tested maintainability by explicitly adding defects to the code and then measured the time taken to remove them. However, modifiability was examined by adding new functionality and then measuring the LOC metric and the time taken to implement these features. The authors did not find a clear effect of refactoring on these two external attributes. 

Stroggylos and Spinellis \citep{stroggylos2007refactoring} opted for
searching words stemming from the verb ``refactor" such
as \say{refactoring} or \say{refactored} to identify commits related to refactoring to study the impact of refactoring on quality using eight object-oriented metrics. Their results indicated possible negative effects of refactoring on quality, \textit{e.g.,} increased LCOM metric. Moser \etal \citep{moser2007case} investigated the impact of refactoring on productivity within an agile team. Their results indicated that refactoring not only enhanced software developers' productivity but also positively affected various quality aspects, such as maintainability. Alshayeb \citep{alshayeb2009empirical} conducted a study aiming to assess the impact of eight refactorings on five external quality attributes (\textit{i.e.,} adaptability, maintainability, understandability, reusability, and testability). The author found that refactoring could improve the quality in some classes, but could also decrease software quality to some extent in other classes. Hegedus \etal \citep{hegedHus2010effect} examined the effect of singular refactoring techniques on testability, error proneness, and other maintainability attributes. They concluded that refactoring could have undesirable side effects that can degrade the quality of the source code. 

Shatnawi and Li \citep{shatnawi2011empirical} used the hierarchical quality model to assess the impact of refactoring on four quality factors in software, namely reusability, flexibility, extendibility, and effectiveness. The authors found that most of the refactoring operations have a positive impact on quality; however, some operations deteriorated quality. Bavota \etal empirically investigated the developers' perception of coupling, as captured by structural, dynamic, semantic, and logical coupling measures. They found that the semantic coupling measure aligns with developers' perceptions better than the other. Bavota \etal \citep{bavota2015experimental} used \texttt{RefFinder}\footnote{https://github.com/SEAL-UCLA/Ref-Finder}, 
a version-based refactoring detection tool, to mine the evolution history of three open-source systems. They mainly investigated the relationship between refactoring and quality. The findings of the study indicate that 42\% of the refactorings performed are affected by code smells, and the refactorings were able to eliminate code smells only in 7\% of the cases. 

Cedrim \etal \citep{cedrim2016does} conducted a longitudinal study involving 25 projects to explore the improvement of the structural quality of software. They examined the relationship between refactorings and code smells, categorizing refactorings based on whether they added or removed problematic code structures. The study results indicate that only 2.24\% of the refactorings removed the code smells, while 2.66\% introduced new ones. Chavez \etal \citep{chavez2017does} investigated the impact of refactoring on five internal quality attributes—cohesion, coupling, complexity, inheritance, and size—using 25 quality metrics. Their study indicated that root-canal refactoring-related operations either improved or at least did not worsen internal quality attributes. Furthermore, when floss refactoring-related operations are applied, 55\% of these operations improved these attributes, while only 10\% quality decreased. Pantiuchina \etal \citep{pantiuchina2020developers} investigated the motivation behind refactoring by computing 42 product and process metrics for each of the 213,102 commits in
the studied projects.

Two studies, particularly relevant to our work, have delved into comment commits in which developers explicitly aimed to improve software quality \citep{szoke2014bulk,pantiuchina2018improving}. Szoke \etal \citep{szoke2014bulk} studied 198 refactoring commits of five large-scale industrial systems to investigate the effects of these commits on the quality of several revisions over a period of time. To understand the purpose of the applied refactorings, developers were trained and asked to articulate the reason when committing changes to repositories, relating to (1) fixing coding issues, (2) addressing anti-patterns, and (3) improving specific metrics. The results of the study showed that performing a single refactoring could negatively impact the quality, but applying refactorings in blocks (\eg fixing more coding issues or improving more quality metrics) can significantly improve software quality. In a related study, Pantiuchina \etal \citep{pantiuchina2018improving} empirically investigated the correlation between seven code metrics and the quality improvement explicitly reported by developers in 1,282 commit messages. The study showed that quality metrics sometimes do not capture the quality improvement reported by developers. Both studies used quality metrics as a common indicator to assess quality improvements, concluding that minor refactoring changes rarely had a substantial impact on software quality.

All of the aforementioned studies have focused on evaluating the impact of refactorings on quality by examining either internal or external quality attributes through various methodologies. Among them, few studies \citep{ratzinger2005improving, stroggylos2007refactoring, szoke2014bulk, bavota2015experimental, cedrim2016does, chavez2017does, pantiuchina2018improving, hamdi2021empirical,ouni2023impact,alomar2020developers,alomar2022refactoring,fakhoury2019improving,alomar2019impact} mined software repositories to explore the impact on quality. Otherwise, the vast majority of these studies 
used a limited set of projects and mined general commits without applying any form of verification regarding whether refactorings have actually been applied. 

Our work differs from these studies shown \textcolor{black}{in Tables \ref{Table:Quality Metrics in Related Work} and \ref{Table:Quality Metrics in Related Work-v2}}, as our main purpose is to explore whether there is an alignment between quality metrics and the removal of code duplication that developers document in the commit messages. As we summarize these state-of-the-art studies, we identify 5 popular quality attributes, namely \textit{Cohesion}, \textit{Coupling}, \textit{Complexity}, \textit{Inheritance}, and \textit{Design size}. Given the varied metrics advocated by different studies to calculate these quality attributes, we extracted and calculated 32 structural metrics. In a more qualitative sense, we conducted an empirical study using 322 distinct commits that are proven to contain real-world instances of refactoring activities, with the purpose of removing code duplication. To the best of our knowledge, no previous study has empirically investigated, using a curated set of commits, the representativeness of structural design metrics for code duplication. The next section details the steps we took to design our empirical setup.
\section{Study Design}
\label{Section:methodology}

\begin{figure*}[t]
\centering 
\includegraphics[width=1.0\textwidth]{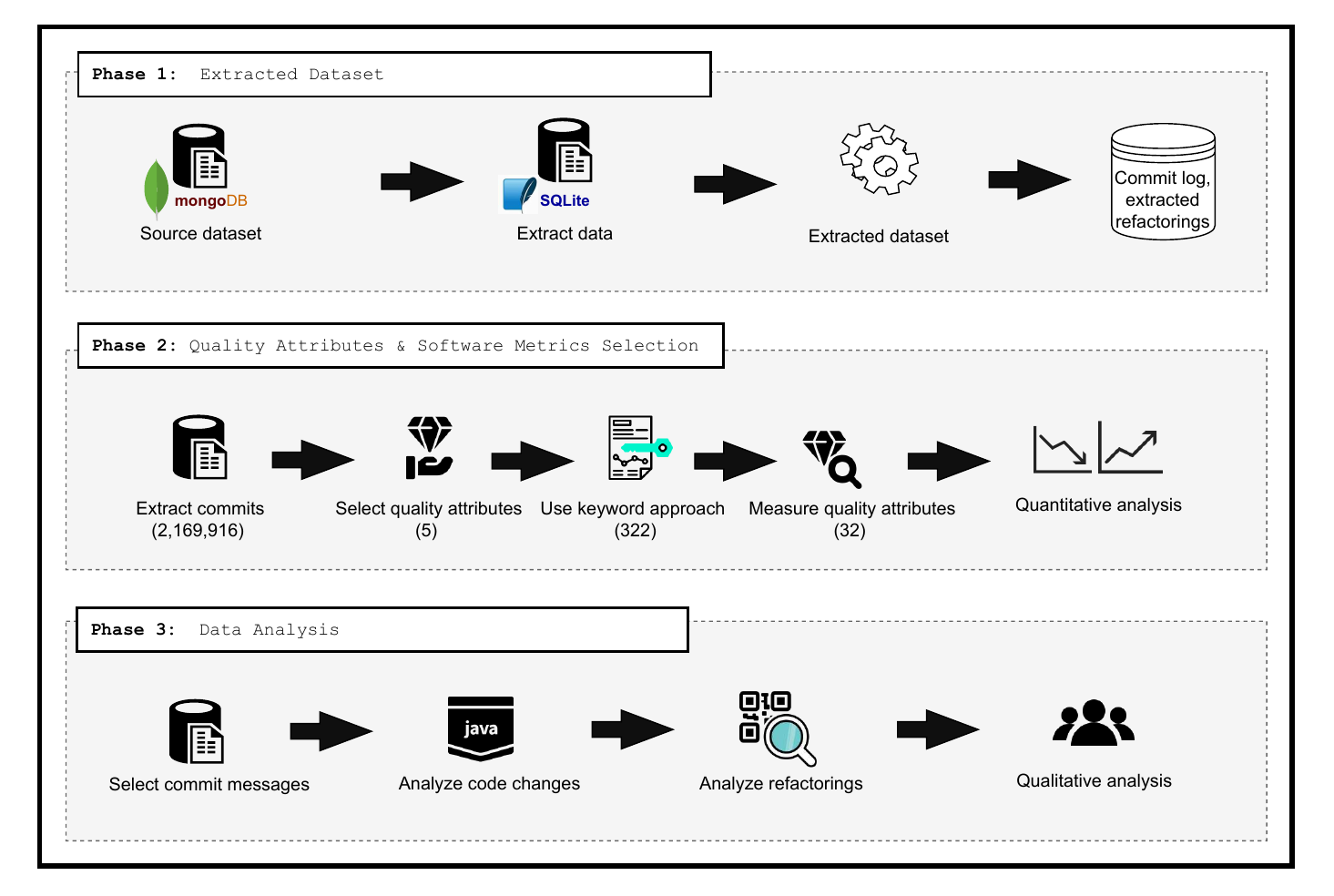}
\caption{\textcolor{black}{Overview of the empirical study design, highlighting the 3 main phases: Dataset Extraction, Selection of Quality Attributes and Software Metrics, and Data Analysis.}}
\label{fig:approach_overview}
\end{figure*}

Our primary objective is to explore the alignment between developers' perceptions of code duplicate removal (as anticipated by developers) and the actual improvement in software quality (as evaluated by quality metrics). Specifically, our aim is to address the following research questions.
\begin{boxK}
\textbf{RQ$_1$}: \textcolor{black}{What is the quantitative code quality assessment of code duplications that have been intentionally removed by developers?}

\textbf{RQ$_2$}: \textcolor{black}{What are the refactoring operations associated with code duplicate removal?}
\end{boxK}

To address our research questions, we conducted a three-phase empirical study. \textcolor{black}{An overview of the experiment methodology is depicted in Figure \ref{fig:approach_overview}. The initial phase involves extracting a substantial number of open-source Java projects along with their instances of refactoring throughout their development history, specifically focusing on commit-level code changes for each project under consideration. In the second phase, we select software quality metrics to compare their values before and after the identified refactoring commits. Subsequently, the third phase involves analyzing commit messages to identify refactoring commits where developers document their perception of code duplicate removal. In the next subsection, we discuss each phase in detail.}

\subsection{Extracted Dataset}

Our study uses the SmartSHARK MongoDB Release 2.2 dataset  \citep{trautsch2021msr}. This dataset contains a wide range of information for 128 open-source Java projects, such as commit history, issues, refactorings, code metrics, mailing lists, and continuous integration data. All Java projects are part of the Apache ecosystem and utilize GitHub as their version control repository and JIRA for issue tracking. SmartSHARK utilizes \texttt{RefDiff} \citep{silva2017refdiff} and \texttt{RefactoringMiner} \citep{tsantalis2018accurate} to mine refactoring operations. \textcolor{black}{This study is motivated to investigate code duplication-aware refactoring practices in Apache projects. A recent study \citep{xiao2024empirical} highlights the Apache Software Foundation as a prominent example of successful open-source software communities \citep{mockus2002two,mockus2000case,crowston2006assessing}. Both practitioners and researchers have been extracting valuable insights and gaining experience from Apache's effective practices to drive the open-source movement forward \citep{rigby2008open,duenas2007apache,weiss2006evolution}. Furthermore, Apache is a collaborative environment where engineers from major corporations such as IBM, Google, Yahoo, Sun, and Oracle volunteer to develop open-source software infrastructure \citep{severance2012apache}.} 


To extract the relevant information, we built custom scripts to extract data pertinent to our study (\ie commits, metrics, refactorings) from
the source dataset into an SQLite database for analysis. First, we extract all commits with the keyword `duplicat*' and `code clone', discussed later in Section \ref{dataanslysis}. Next,
we extract all refactoring operations. However, due to the use of two refactoring mining tools, there are duplicate operations in the source data. \textcolor{black}{Hence, our next step is to remove all duplicates by comparing the refactoring descriptions. After that, we select all
commits associated with a refactoring operation. Using both refactoring mining tools allowed us to mitigate the limitations of relying on a single tool and ensured a more diverse and thorough dataset.} Table \ref{Table:DATA_Overview} summarizes the extracted data.

\subsection{Quality Attributes \& Quality Metrics Selection}
To setup a comprehensive set of quality attributes for evaluation in our study, we initially analyze existing studies to identify commonly recognized software quality attributes \citep{chidamber1994metrics,lorenz1994object,mccabe1976complexity, henry1981software, nejmeh1988npath, Destefanis:2014:SMA:2813544.2813555}. Next, we assess whether the metrics evaluate various object-oriented design aspects, mapping each internal quality attribute to the corresponding structural metric(s). 
  Additionally, we extract associations between metrics (such as the CK suite \citep{chidamber1994metrics}, McCabe \citep{mccabe1976complexity}, and Lorenz and Kidd's book \citep{lorenz1994object}) and internal quality attributes from the literature review. \textcolor{black}{Tables \ref{Table:Quality Metrics in Related Work} and \ref{Table:Quality Metrics in Related Work-v2} summarize the extracted metrics.}

Subsequently, we examined the extracted metrics to determine whether these metrics exist in the SmartSHARK dataset, calculated using OpenStaticAnalyzer\footnote{https://github.com/sed-inf-u-szeged/OpenStaticAnalyzer}. The extraction process results in 32 distinct structural metrics as shown in Table \ref{Table:Quality Metrics Used in This Study.}. The list of metrics is (1) well-known and defined in the literature, and (2) can assess different code-level elements, \ie method, class, package. 


\subsection{Data Analysis}
\label{dataanslysis}

\begin{table}[h!]
\begin{center}
\caption{\textcolor{black}{Summary of the extracted data.}}
\label{Table:DATA_Overview}
\begin{adjustbox}{width=1.0\textwidth,center}
\begin{tabular}{lllll}\hline
\toprule
\bfseries Item & \bfseries Count \\
\midrule
Total projects & 128 \\
\cellcolor{gray!30}Refactoring commits with keyword `\textit{duplicat*}' & \cellcolor{gray!30}2,169,916  \\
False positive commits & 22 \\
\cellcolor{gray!30}Refactoring commits after removing false positives & \cellcolor{gray!30}2,164,797 \\
(Distinct) Refactoring commits with keyword `\textit{duplicat*}' & 332  \\
\bottomrule
\end{tabular}
\end{adjustbox}
\end{center}
\end{table}

After extracting all refactoring commits, we want to keep only commits where refactoring is documented. We continue to filter them, using the content of their messages at this stage. We use a keyword-based search to find commits whose messages contain the keywords (\ie `duplicat*' or `code clone*'). We selected these keywords because these keywords are naturally used by developers to articulate their intent regarding code duplication \citep{alomar2019can,alomar2021we}. However, it is worth mentioning that we did not find any commits with the keyword `code clone'. Therefore, all the commits in our dataset solely include the keyword `duplicat'.

This keyword-based filtering selected 2,169,916 commit messages. 
To ensure that these commits reported developers' intention to remove code duplication, we manually inspected and read through 322 distinct refactoring commits to remove false positives. An example of a discarded commit is: \say{\textit{DeferredDuplicates.java}}. We discarded this commit because the keyword `duplicat' is actually part of the identifier name of the class. In the case of doubts about including a certain commit, it was excluded. This step resulted in considering 322 commits. Our goal is to have a \textit{gold set} of commits in which the developers explicitly reported the removal of duplicate code. This \textit{gold set} will serve to check later if there is an alignment between the real quality metrics affected in the source code, and the code duplicate removal as documented by developers. 
 An example of commit messages belonging to the \textit{gold set}, is showcased in the following commit message  \say{\textit{Refactored JavaClass and FieldOrMethod to avoid a code duplication}}. 

\begin{table}
  \centering
	 \caption{\textcolor{black}{Structural code quality metrics used in this study.}}
	 \label{Table:Quality Metrics Used in This Study.}
  \begin{threeparttable}
\begin{adjustbox}{width=1.0\textwidth,center}
\begin{tabular}{llll}\hline
\toprule
\bfseries Quality Attribute & \bfseries Study &   \bfseries Metric & \bfseries Description \\
\midrule
Cohesion & \cite{pantiuchina2018improving,chavez2017does} &↓ LCOM& Lack of Cohesion of Methods   \\ 
Coupling &  \cellcolor{gray!30}\cite{chavez2017does,pantiuchina2018improving} & \cellcolor{gray!30}↓ CBO&\cellcolor{gray!30}Coupling Between Objects    \\
         & \cite{pantiuchina2018improving} & ↓ RFC & Response For Class   \\
         & \cellcolor{gray!30}\cite{islam2018characteristics} & \cellcolor{gray!30}↓ NII &\cellcolor{gray!30}Number of Incoming Invocations  \\
         & \cite{islam2018characteristics} & ↓ NOI &Number of Outgoing Invocations \\
Complexity & \cellcolor{gray!30}\cite{chavez2017does} & \cellcolor{gray!30}↓ CC & \cellcolor{gray!30}Cyclomatic Complexity 
           \\
           & \cite{chavez2017does,pantiuchina2018improving,singh2012evaluation} & ↓ WMC& Weighted Method Count  \\
           & \cellcolor{gray!30}\cite{islam2018characteristics} & \cellcolor{gray!30}↓ NL & \cellcolor{gray!30}Nesting Level  \\
           & \cite{islam2018characteristics} & ↓ NLE &Nesting Level Else-if  \\
           & \cellcolor{gray!30}\cite{islam2018characteristics} & \cellcolor{gray!30}↓ HCPL & \cellcolor{gray!30}Hal. Calculated Program Length \\
            & \cite{islam2018characteristics} & ↓ HDIF & Hal. Difficulty  \\
             & \cellcolor{gray!30}\cite{islam2018characteristics} & \cellcolor{gray!30}↓ HEFF & \cellcolor{gray!30}Hal. Effort  \\
              & \cite{islam2018characteristics} & ↓ HNDB & Hal. Number of Delivered Bugs   \\
               & \cellcolor{gray!30}\cite{islam2018characteristics} &\cellcolor{gray!30}↓ HPL & \cellcolor{gray!30}Hal. Program Length   \\
                & \cite{islam2018characteristics} & ↓ HPV & Hal. Program Vocabulary  \\
                 & \cellcolor{gray!30}\cite{islam2018characteristics} & \cellcolor{gray!30}↓ HTRP &\cellcolor{gray!30}Hal. Time Required to Program  \\
                  & \cite{islam2018characteristics} & ↓ HVOL &Hal. Volume \\
                   & \cellcolor{gray!30}\cite{islam2018characteristics} &\cellcolor{gray!30}↑ MIMS & \cellcolor{gray!30}Maintainability Index (MS) \\
                    & \cite{islam2018characteristics} &↑ MI& Maintainability Index (OV) \\
                     & \cellcolor{gray!30}\cite{islam2018characteristics} &\cellcolor{gray!30}↑ MISEI& \cellcolor{gray!30}Maintainability Index (SEIV) \\
                      & \cite{islam2018characteristics} &↑ MISM&  Maintainability Index (SV)\\
Inheritance & \cellcolor{gray!30}\cite{chavez2017does,singh2012evaluation} & \cellcolor{gray!30}↓ DIT &\cellcolor{gray!30}Depth of Inheritance Tree 
  \\
   & \cite{chavez2017does,singh2012evaluation} & ↓ NOC &Number of Children   \\
 & \cellcolor{gray!30}\cite{bavota2015experimental} & \cellcolor{gray!30}↓ NOA & \cellcolor{gray!30}Number of Operations Added by Subclass   \\
        
Design Size & \cite{chavez2017does} & ↓ LOC & Lines of Code \\
& \cellcolor{gray!30}\cite{islam2018characteristics} &\cellcolor{gray!30}↓ TLOC &\cellcolor{gray!30}Total Lines of Code   \\
& \cite{chavez2017does} & ↓ LLOC &Logical Lines of Code   \\
& \cellcolor{gray!30}\cite{islam2018characteristics} & \cellcolor{gray!30}↓ TLLOC&\cellcolor{gray!30}Total Logical Lines of Code  \\
            & \cite{chavez2017does} & ↑  CLOC&Lines with Comments  \\
            & \cellcolor{gray!30}\cite{stroggylos2007refactoring} & \cellcolor{gray!30}↓ NPM &\cellcolor{gray!30}Number of Public Methods  \\
            &\cite{islam2018characteristics} &↓ NOS& Number of Statements  \\
            &\cellcolor{gray!30}\cite{islam2018characteristics} & \cellcolor{gray!30}↓ TNOS&\cellcolor{gray!30}Total Number of Statements  \\
\bottomrule
\multicolumn{4}{l}{\tiny 
↑ by a metric indicates the higher the better for that metric; 
↓ by a metric indicates the lower the better for that metric.}
\end{tabular}
\end{adjustbox}
\end{threeparttable}
\end{table}

We perform a qualitative analysis of intriguing instances of alignment or disparity between the removal of code duplication as perceived by developers and its evaluation through quality metrics. To do this, the author manually inspects the commits, which involves analyzing the diff code alongside the metrics profile of the affected code elements before and after the commit.

The resulting commits correspond to our data points, each data point is represented by a set of \textit{pre-refactoring} and \textit{post-refactoring} Java files. These data points will be used in the experiments, to measure the effect of changes in terms of structural metrics, with respect to the quality attribute, announced in the commit message.

\section{Results \& Discussion}
\label{Section:Result}
\subsection{\textcolor{black}{What is the quantitative code quality assessment of code duplications that have been intentionally removed by developers?}}
For each refactoring commit in which developers document the removal of duplicate code, we extract its associated metric values  (see Table~\ref{Table:Quality Metrics Used in This Study.}) before and after the commit. 
 In other words, for commit messages related to the removal of code duplicates, we examine 32 corresponding metric values before and after the selected refactoring commit. As we evaluate metric values both pre- and post-refactoring, we want to distinguish, for each metric, whether there is a variation between its pair of values, whether this variation signifies an improvement, and whether the variation is statistically significant. Therefore, we use the Wilcoxon test \citep{wilcoxon1945individual}, a non-parametric test, to compare the group of metric values before and after the commit since these groups depend on each other. The null hypothesis is defined by no variation in the metric values of pre- and post-refactored code elements. Thus, the alternative hypothesis indicates a variation in the metric values. In each case, a decrease in the metric value is considered desirable (\ie an improvement), except for complexity metrics related to the maintainability index and CLOC (see Table~\ref{Table:Quality Metrics Used in This Study.}), where higher values are desirable. Furthermore, the variation between the values of both sets is considered significant if its associated \textit{p}-value is less than 0.05. Furthermore, we used the Cliff's delta ($\delta$) effect size to estimate the magnitude of the differences. Regarding its interpretation, we follow the guidelines reported by Grissom \etal \citep{trove.nla.gov.au/work/16432558}:

 \begin{itemize}
\item Negligible for $\mid \delta \mid< 0.147$
\item Small for $0.147 \leq \mid \delta \mid < 0.33$
\item Medium for $0.33 \leq \mid \delta \mid < 0.474$
\item Large for $\mid \delta \mid \geq 0.474$
\end{itemize}

\begin{figure*}
\centering
\centering\includegraphics[width=3.5cm]{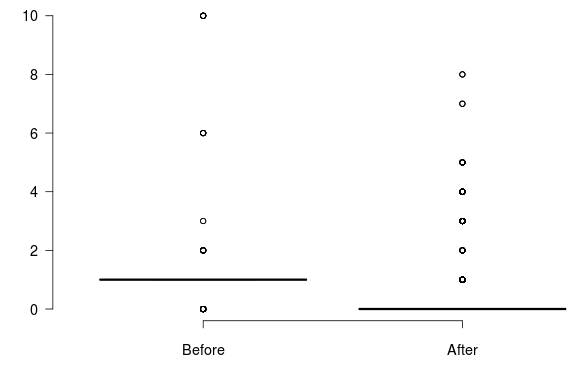}
\caption{Cohesion - LCOM5}
\label{BP:chesion-lcom}
\caption{\textcolor{black}{Boxplots of cohesion metric values of pre- and post-refactored files.}}
\label{Chart:Boxplots_cohesion}
\end{figure*}
\begin{figure*}
\centering
\begin{subfigure}{3.5cm}
\centering\includegraphics[width=3.5cm]{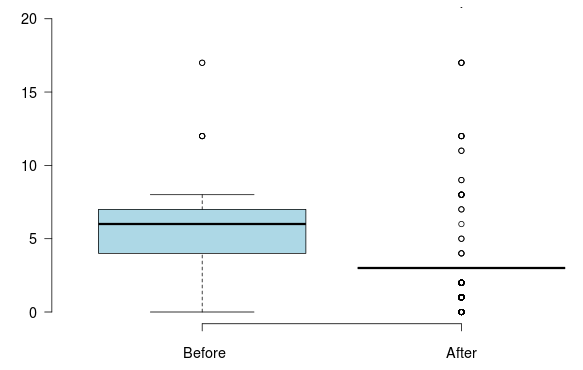}
\caption{Coupling - CBO}
\label{BP:coupling-cbo}
\end{subfigure}%
\begin{subfigure}{3.5cm}
\centering\includegraphics[width=3.5cm]{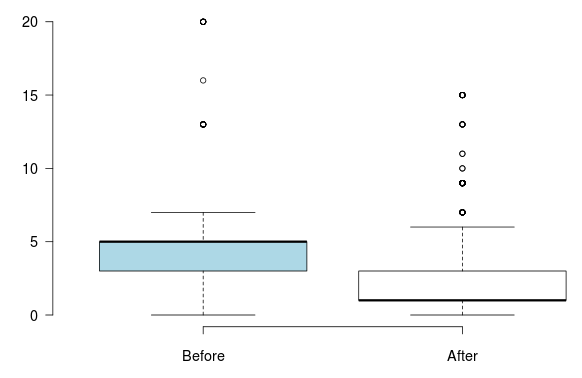}
\caption{Coupling - RFC}
\label{BP:coupling-rfc}
\end{subfigure}
\begin{subfigure}{3.5cm}
\centering\includegraphics[width=3.5cm]{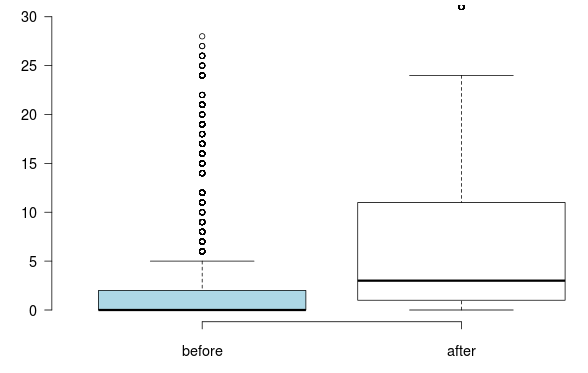}
\caption{Coupling - NII}
\label{BP:coupling-nii}
\end{subfigure}
\begin{subfigure}{3.5cm}
\centering\includegraphics[width=3.5cm]{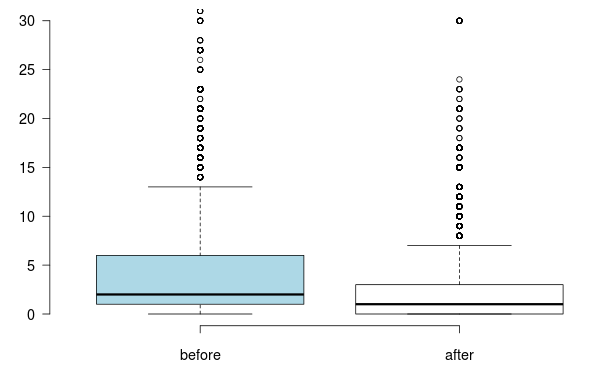}
\caption{Coupling - NOI}
\label{BP:coupling-noi}
\end{subfigure}%
\caption{\textcolor{black}{Boxplots of coupling metric values of pre- and post-refactored files.}}
\label{Chart:Boxplots_coupling}
\end{figure*}
\begin{figure*}
\centering
\begin{subfigure}{3.5cm}
\centering\includegraphics[width=3.5cm]{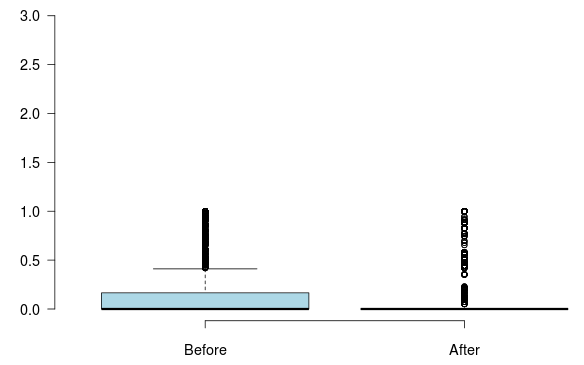}
\caption{Complexity - CC}
\label{BP:Complexity-cc}
\end{subfigure}%
\begin{subfigure}{3.5cm}
\centering\includegraphics[width=3.5cm]{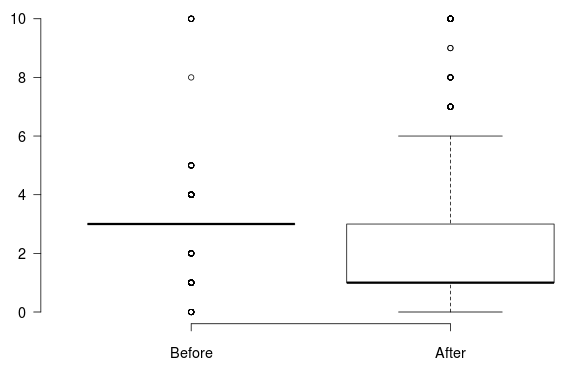}
\caption{Complexity - WMC}
\label{BP:Complexity-wmc}
\end{subfigure}%
\begin{subfigure}{3.5cm}
\centering\includegraphics[width=3.5cm]{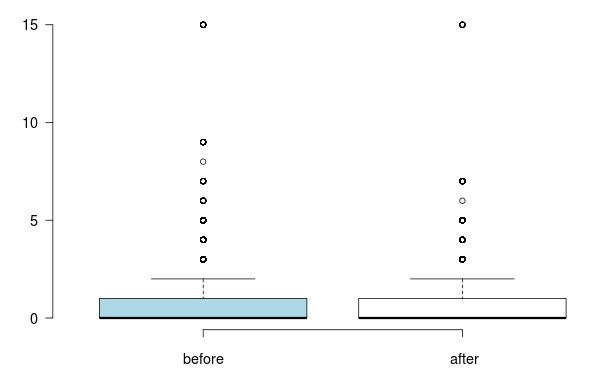}
\caption{Complexity - NL}
\label{BP:Complexity-nl}
\end{subfigure}%
\begin{subfigure}{3.5cm}
\centering\includegraphics[width=3.5cm]{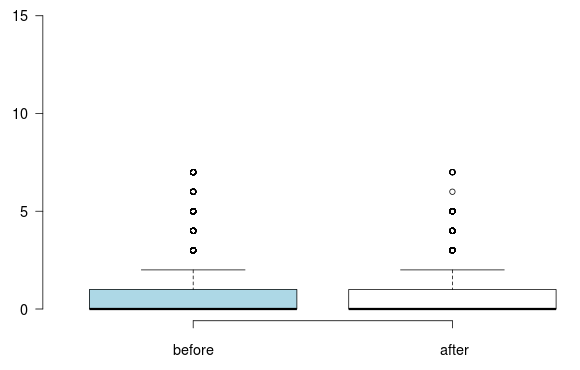}
\caption{Coupling - NLE}
\label{BP:coupling-nle}
\end{subfigure}%
\vspace{9pt}
\begin{subfigure}{3.5cm}
\centering\includegraphics[width=3.5cm]{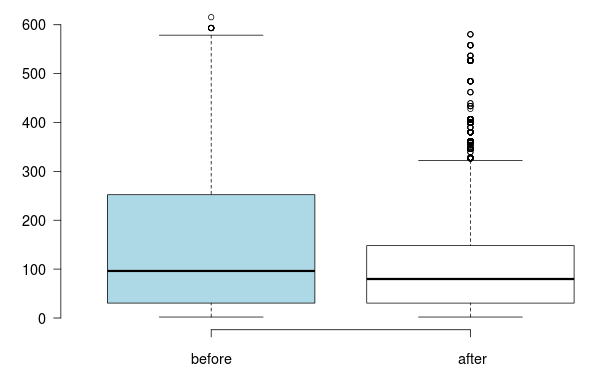}
\caption{Complexity - HCPL}
\label{BP:Complexity-hcpl}
\end{subfigure}%
\begin{subfigure}{3.5cm}
\centering\includegraphics[width=3.5cm]{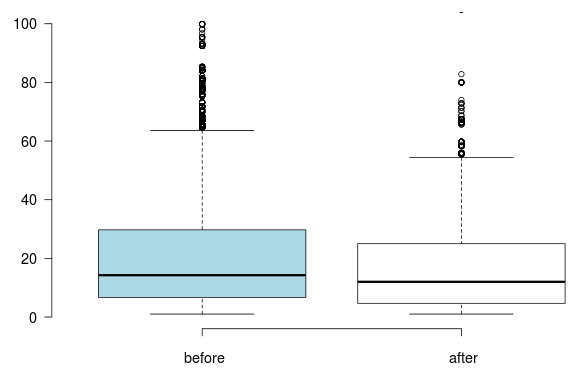}
\caption{Complexity - HDIF}
\label{BP:Complexity-hdif}
\end{subfigure}%
\begin{subfigure}{3.5cm}
\centering\includegraphics[width=3.5cm]{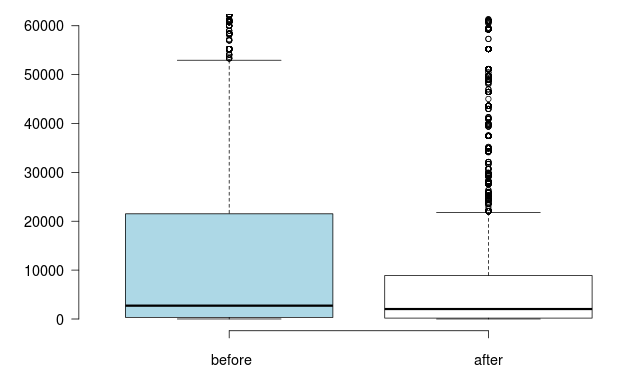}
\caption{Complexity - HEFF}
\label{BP:Complexityheff}
\end{subfigure}%
\begin{subfigure}{3.5cm}
\centering\includegraphics[width=3.5cm]{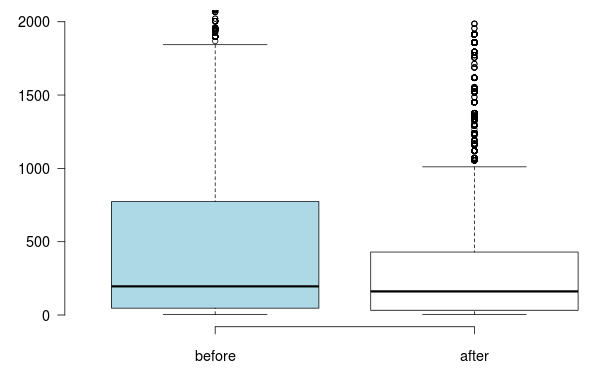}
\caption{Complexity - HNDB}
\label{BP:Complexity-hndb}
\end{subfigure}%
\vspace{9pt}
\begin{subfigure}{3.5cm}
\centering\includegraphics[width=3.5cm]{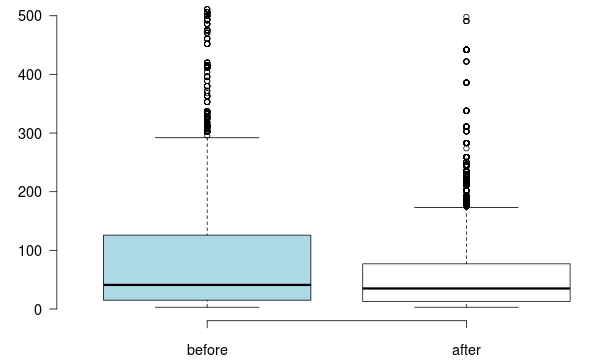}
\caption{Complexity - HPL}
\label{BP:Complexity-hpl}
\end{subfigure}%
\begin{subfigure}{3.5cm}
\centering\includegraphics[width=3.5cm]{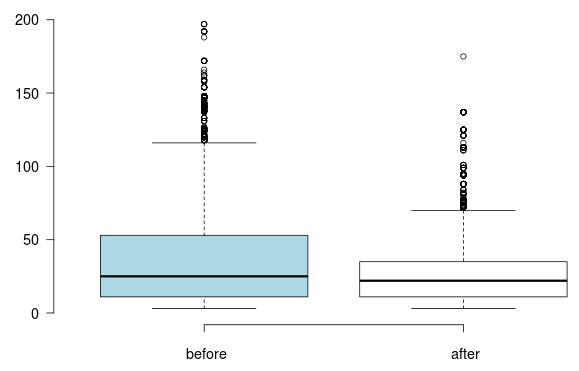}
\caption{Complexity - HPV}
\label{BP:Complexity-hpv}
\end{subfigure}%
\begin{subfigure}{3.5cm}
\centering\includegraphics[width=3.5cm]{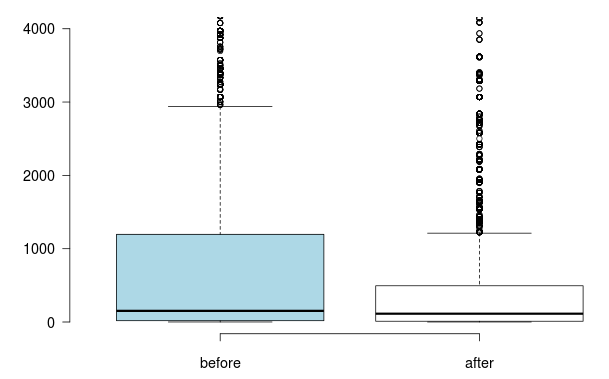}
\caption{Complexity - HTRP}
\label{BP:Complexity-htrp}
\end{subfigure}%
\begin{subfigure}{3.5cm}
\centering\includegraphics[width=3.5cm]{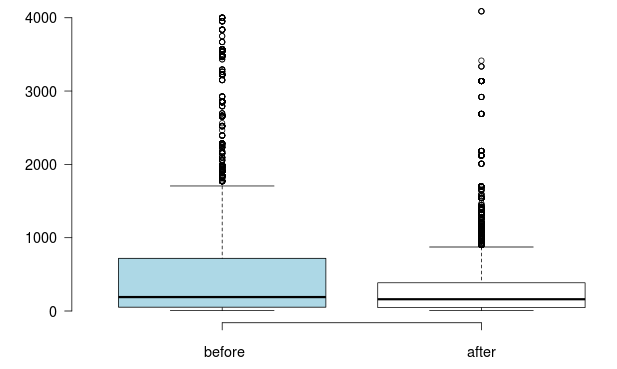}
\caption{Complexity - HVOL}
\label{BP:Complexity-hvol}
\end{subfigure}%
\vspace{9pt}
\begin{subfigure}{3.5cm}
\centering\includegraphics[width=3.5cm]{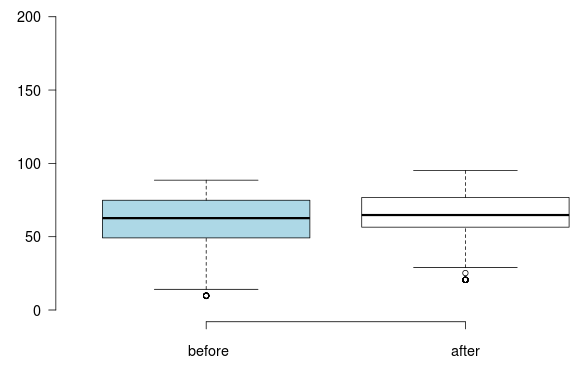}
\caption{Complexity - MIMS}
\label{BP:Complexity-mims}
\end{subfigure}%
\begin{subfigure}{3.5cm}
\centering\includegraphics[width=3.5cm]{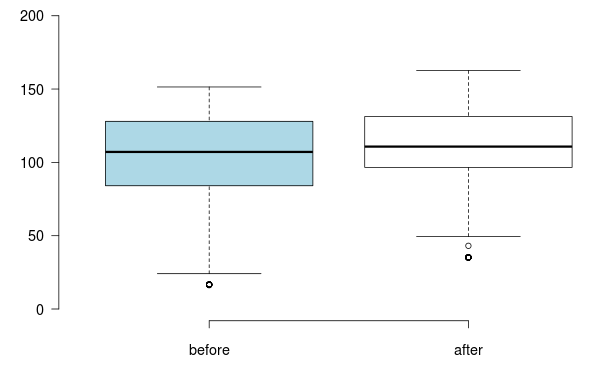}
\caption{Complexity - MI}
\label{BP:Complexity-mi}
\end{subfigure}%
\begin{subfigure}{3.5cm}
\centering\includegraphics[width=3.5cm]{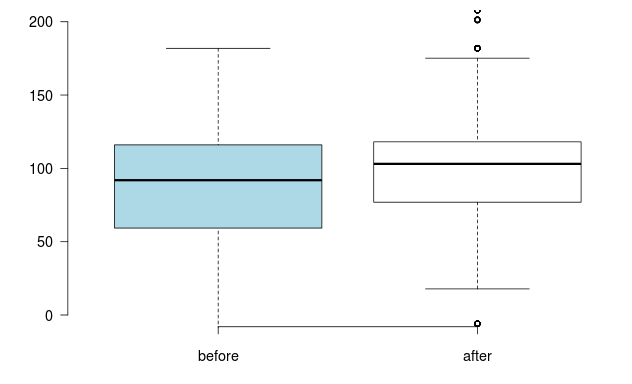}
\caption{Complexity - MISEI}
\label{BP:Complexity-misei}
\end{subfigure}%
\begin{subfigure}{3.5cm}
\centering\includegraphics[width=3.5cm]{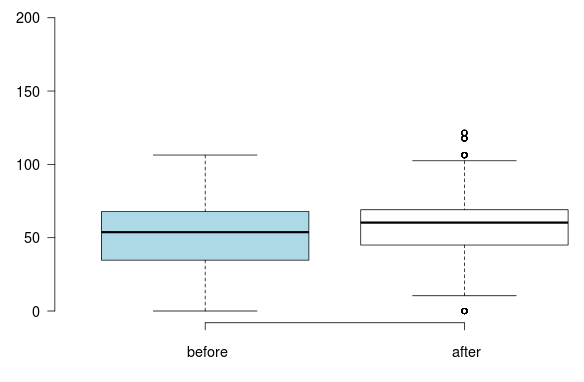}
\caption{Complexity - MISM}
\label{BP:Complexity-mism}
\end{subfigure}%
\caption{\textcolor{black}{Boxplots of complexity metric values of pre- and post-refactored files.}}
\label{Chart:Boxplots_complexity}
\end{figure*}
\begin{figure*}
\centering
\begin{subfigure}{3.5cm}
\centering\includegraphics[width=3.5cm]{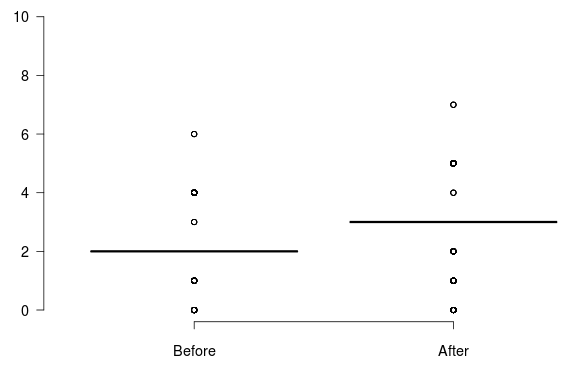}
\caption{Inheritance - DIT}
\label{BP:Inheritance-dit}
\end{subfigure}
\begin{subfigure}{3.5cm}
\centering\includegraphics[width=3.5cm]{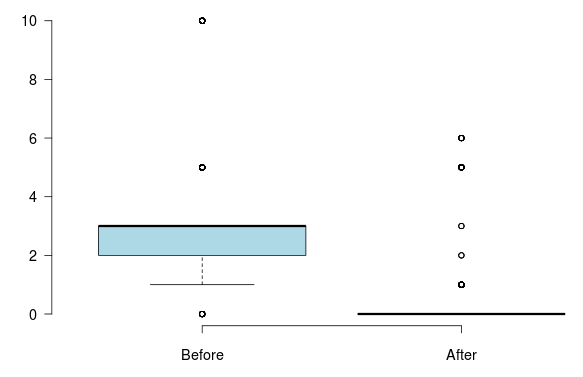}
\caption{Inheritance - NOC}
\label{BP:Inheritance-noc}
\end{subfigure}%
\begin{subfigure}{3.5cm}
\centering\includegraphics[width=3.5cm]{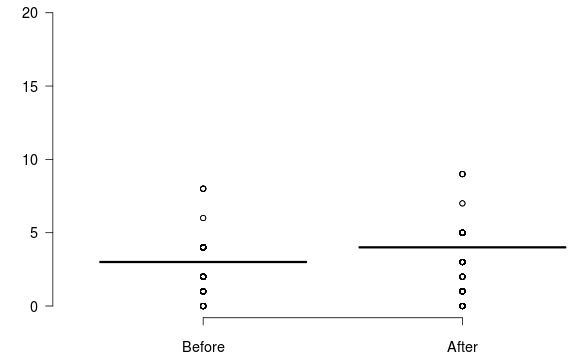}
\caption{Inheritance - NOA}
\label{BP:Inheritance-noa}
\end{subfigure}%
\caption{\textcolor{black}{Boxplots of inheritance metric values of pre- and post-refactored files.}}
\label{Chart:Boxplots_inheritance}
\end{figure*}
\begin{figure*}
\centering
\begin{subfigure}{3.5cm}
\centering\includegraphics[width=3.5cm]{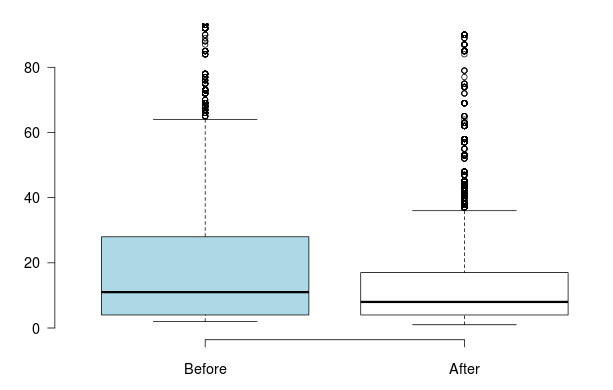}
\caption{Design Size - LOC}
\label{BP:Design Size-loc}
\end{subfigure}%
\begin{subfigure}{3.5cm}
\centering\includegraphics[width=3.5cm]{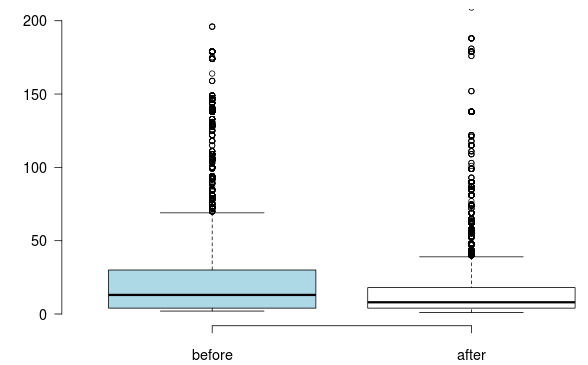}
\caption{Design Size - TLOC}
\label{BP:Design Size-tloc}
\end{subfigure}%
\begin{subfigure}{3.5cm}
\centering\includegraphics[width=3.5cm]{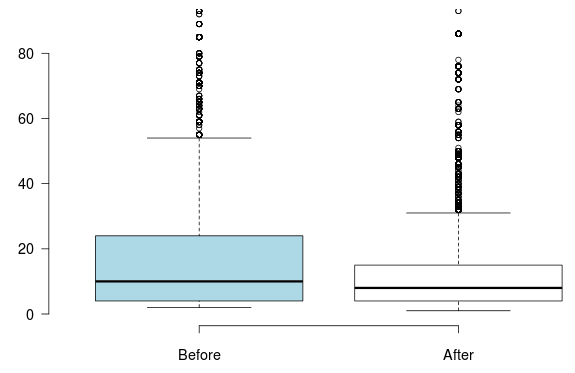}
\caption{Design Size - LLOC}
\label{BP:Design Size-lloc}
\end{subfigure}%
\begin{subfigure}{3.5cm}
\centering\includegraphics[width=3.5cm]{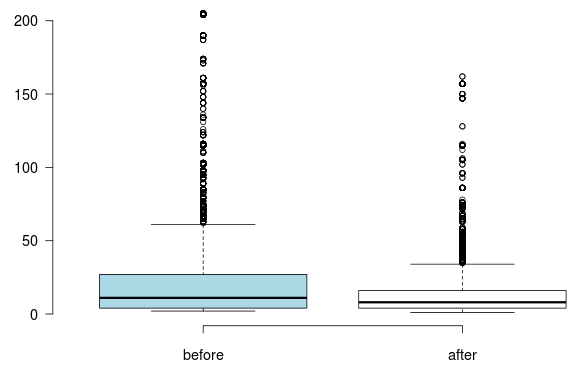}
\caption{Design Size - TLLOC}
\label{BP:Design Size-tlloc}
\end{subfigure}%
\vspace{9pt}
\begin{subfigure}{3.5cm}
\centering\includegraphics[width=3.5cm]{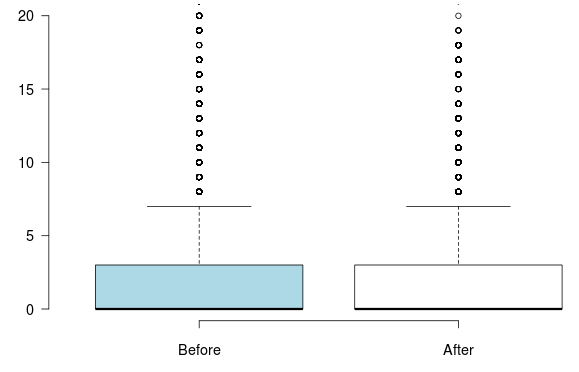}
\caption{Design Size - CLOC}
\label{BP:Design Size-cloc}
\end{subfigure}%
\begin{subfigure}{3.5cm}
\centering\includegraphics[width=3.5cm]{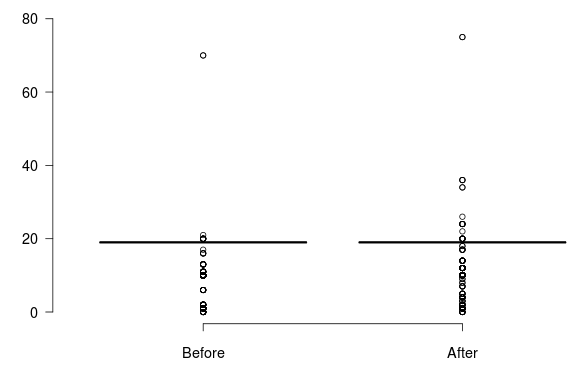}
\caption{Design Size - NPM}
\label{BP:Design Size-npm}
\end{subfigure}%
\begin{subfigure}{3.5cm}
\centering\includegraphics[width=3.5cm]{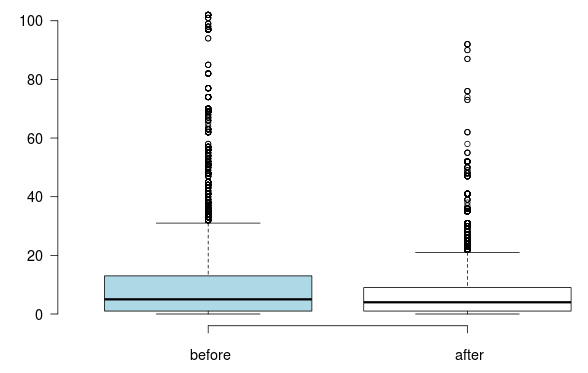}
\caption{Design Size - NOS}
\label{BP:Design Size-nos}
\end{subfigure}%
\begin{subfigure}{3.5cm}
\centering\includegraphics[width=3.5cm]{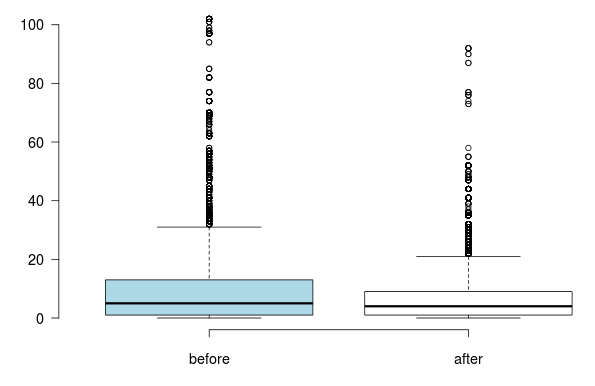}
\caption{Design Size - TNOS}
\label{BP:Design Size-tnos}
\end{subfigure}

\caption{\textcolor{black}{Boxplots of design size metric values of pre- and post-refactored files.}}
\label{Chart:Boxplots_design size}
\end{figure*}

To answer our main research question, we provide a detailed analysis of each of the five quality attributes reported in Table \ref{Table:Quality Metrics Used in This Study.} and qualitatively analyze the cases with positive and negative impacts. Table~\ref{Table:Metrics Suites and Metrics Tools Summary} shows the overall impact of refactorings on quality. The boxplots in \textcolor{black}{Figures \ref{Chart:Boxplots_cohesion}, \ref{Chart:Boxplots_coupling}, \ref{Chart:Boxplots_complexity}, \ref{Chart:Boxplots_inheritance}, and \ref{Chart:Boxplots_design size}} show the distribution of each metric before and after each of the examined commits.

\begin{table}
  \centering
\caption{\textcolor{black}{Effect of duplicate code removal on structural metrics. (+ve) indicates positive impact; (-ve) indicates negative impact; (-) indicates metric remains unaffected, \textbf{bold} indicates statistical significance; \textit{italic} indicates improvement.}}
\label{Table:Metrics Suites and Metrics Tools Summary}
\begin{adjustbox}{width=1.0\textwidth,center}
\begin{tabular}{lllll}\hline
\toprule
\bfseries Quality Attribute & \bfseries Metric & \bfseries Impact & \bfseries \textit{p}-value & \bfseries Cliff's delta ($\delta$) \\
\midrule
Cohesion &  LCOM5  & +ve & \textit{\textbf{7.72e-41}} & 0.54 (Large)
\\ 
Coupling &  \cellcolor{gray!30}CBO  & \cellcolor{gray!30}+ve & \cellcolor{gray!30}\textit{\textbf{9.49e-76}} & \cellcolor{gray!30}0.6 (Large)
\\
         &  RFC & +ve & \textit{\textbf{1.25e-68}}  & 0.55 (Large)

\\
         &  \cellcolor{gray!30}NII & \cellcolor{gray!30}-ve &  \cellcolor{gray!30}\textbf{0} & \cellcolor{gray!30}0.47 (Large)

\\
         &  NOI & +ve & \textit{\textbf{0}}  & 0.26 (Small)

\\
Complexity &  \cellcolor{gray!30}CC & \cellcolor{gray!30}- & \cellcolor{gray!30}\textbf{0} & \cellcolor{gray!30}0.14 (Small)

\\
           &  WMC & +ve & \textit{\textbf{6.51e-70}} & 0.5 (Large)

\\
         &  \cellcolor{gray!30}NL &  \cellcolor{gray!30}- &  \cellcolor{gray!30}\textbf{3.92e-05}&  \cellcolor{gray!30}0.03 (Negligible)

\\
         &  NLE &  - & \textbf{0.004}  & 0.02 (Negligible)

\\
         &  \cellcolor{gray!30}HCPL & \cellcolor{gray!30}+ve &  \cellcolor{gray!30}\textit{\textbf{0}} &  \cellcolor{gray!30}0.14 (Negligible)

\\
         &  HDIF & +ve & \textit{\textbf{0}}  &  0.08 (Negligible)

\\
         &  \cellcolor{gray!30}HEFF & \cellcolor{gray!30}+ve &  \cellcolor{gray!30}\textit{\textbf{2.45e-271}} &  \cellcolor{gray!30}0.13 (Negligible)

\\
         &  HNDB & +ve &  \textit{\textbf{1.07e-266}} & 0.13  (Negligible)

\\
         &  \cellcolor{gray!30}HPL & \cellcolor{gray!30}+ve & \cellcolor{gray!30}\textit{\textbf{0}} & \cellcolor{gray!30}0.13  (Negligible)

\\
         &  HPV & +ve & \textit{\textbf{0}}  & 0.14  (Negligible)

\\
         &  \cellcolor{gray!30}HTRP & \cellcolor{gray!30}+ve &  \cellcolor{gray!30}\textit{\textbf{2.48e-271}} & \cellcolor{gray!30}0.13  (Negligible)

\\
         &  HVOL & +ve & \textit{\textbf{0}}  &  0.13  (Negligible)

\\
         &  \cellcolor{gray!30}MIMS & \cellcolor{gray!30}+ve & \cellcolor{gray!30}\textit{\textbf{7.23e-227}}  &  \cellcolor{gray!30}0.13  (Negligible)

\\
         &   MI & +ve &  \textit{\textbf{7.22e-227}} &  0.13  (Negligible)

\\
         &   \cellcolor{gray!30}MISEI & \cellcolor{gray!30}+ve & \cellcolor{gray!30}\textit{\textbf{0}} & \cellcolor{gray!30}0.16  (Small)

\\
         &   MISM &  +ve&  \textit{\textbf{0}}  & 0.16  (Small)

\\
Inheritance &   \cellcolor{gray!30}DIT & \cellcolor{gray!30}-ve & \cellcolor{gray!30}\textbf{3.81e-199} & \cellcolor{gray!30}0.6 (Large) 
 
\\
            &  NOC & +ve & \textbf{\textit{3.61e-130}} & 0.83 (Large)  

\\
            &  \cellcolor{gray!30}NOA & \cellcolor{gray!30}-ve & \cellcolor{gray!30}\textbf{2.37e-196}  & \cellcolor{gray!30}0.63 (Large)
 
\\ 
Design Size &  LOC & +ve & \textbf{\textit{0}} &   0.14 (Small)

\\
         &  \cellcolor{gray!30}TLOC & \cellcolor{gray!30}+ve &   \cellcolor{gray!30}\textit{\textbf{0}} &  \cellcolor{gray!30}0.16 (Small)

\\
            &  LLOC &  +ve &  \textbf{\textit{0}} &   0.13 (Negligible)

\\
         &  \cellcolor{gray!30}TLLOC & \cellcolor{gray!30}+ve & \cellcolor{gray!30}\textit{\textbf{0}}  &   \cellcolor{gray!30}0.15 (Small)

\\
            &   CLOC & - &  \textbf{1.43e-05} &   0.02 (Negligible)

\\
            &  \cellcolor{gray!30}NPM & \cellcolor{gray!30}- & \cellcolor{gray!30}\textbf{4.42e-193} & \cellcolor{gray!30}0.5 (Large)

\\
         &  NOS &  +ve&  \textit{\textbf{0}} &  0.07 (Negligible)

\\
         &  \cellcolor{gray!30}TNOS & \cellcolor{gray!30}+ve & \cellcolor{gray!30}\textit{\textbf{0}}  &  \cellcolor{gray!30}0.08 (Negligible)

\\
\bottomrule
\end{tabular}
\end{adjustbox}
\end{table}
\noindent\textbf{\textcolor{black}{Cohesion.}}  For commits wherein the messages indicate the removal of code duplication, the boxplot depicted in Figure \ref{BP:chesion-lcom} illustrates the pre- and post-refactoring results of the normalized LCOM. This metric, commonly used in the literature to assess cohesion, is crucial in estimating the strength of cohesion within classes. A lower LCOM metric value generally suggests that classes should be split into one or more classes with better cohesion. Therefore, a low value for this metric signifies strong class cohesiveness. We specifically chose the normalized LCOM metric as it has been widely recognized in the literature  \citep{pantiuchina2018improving,chavez2017does,henderson1995object} as being the alternative to the original LCOM, by addressing its main limitations (artificial outliers, misperception of getters and setters, etc.). As can be seen from the boxplot in Figure \ref{BP:chesion-lcom}, the median drops from 1 to 0. This result indicates that LCOM is improved after code duplicate removal. Furthermore, as shown in Table~\ref{Table:Metrics Suites and Metrics Tools Summary}, LCOM has a positive impact on cohesion quality, as it decreases in the refactored code. This implies that developers did improve the cohesion of their classes. Table~\ref{Table:Metrics Suites and Metrics Tools Summary} shows that the differences in LCOM are statistically significant and the magnitude of the differences is large.

\noindent\textbf{Example (Positive Impact):} To illustrate an improvement in cohesion when the removal of duplicate code was found in the Maven project\footnote{\textcolor{black}{\url{https://github.com/apache/maven-surefire/commit/d5de47a4f790ea2d18edb5e05c1ef2adcd2db8a2}}}, developers applied `Move Class' refactoring to move \texttt{JUnit4TestCheckerTest.MySuite2} to \texttt{JUnit3TestCheckerTest.MySuite2}. This results in its LCOM5 dropping from 2 to 1. The improvement in the LCOM5 metric after the removal of code duplication could be attributed to the simplification of method interactions, the modularization of logic, the enhancement of code clarity, and the abstraction of common functionality.

\noindent\textbf{Example (Negative Impact):} To illustrate a decrease in cohesion when the removal of duplicated code was found in the Maven project\footnote{\textcolor{black}{\url{https://github.com/apache/maven-surefire/commit/d5de47a4f790ea2d18edb5e05c1ef2adcd2db8a2}}}, developers applied `Move Class' refactoring to move \texttt{JUnit4TestCheckerTest.NestedTC} to \texttt{JUnit3TestChecker\break Test.NestedTC}. This results in its LCOM5 increasing from 0 to 2. The LCOM5 metric might not have improved after removing duplicated code, as removal of duplicated code might not have substantially altered the underlying design and interactions of methods.

\noindent\textbf{\textcolor{black}{Coupling.}} For commits with messages indicating the removal of code duplication, the boxplots presented in Figures \ref{BP:coupling-cbo}, \ref{BP:coupling-rfc}, \ref{BP:coupling-nii}, and \ref{BP:coupling-noi} show the pre- and post-refactoring results of four structural metrics, \ie CBO, RFC, NII, and NOI, used in the literature to estimate the coupling. The figures reveal that three of the coupling metrics exhibited an improvement in median values. For instance, CBO, RFC, and NOI medians decreased, respectively, from 6 to 3, from 5 to 1, and from 6 to 3, respectively. CBO counts the number of classes coupled to a particular class through method or attribute calls. Calls are counted in both directions. CBO values have significantly decreased, making it a good coupling representative.  The RFC, which measures the visibility of a class to outsider classes, has been reduced as developers intend to optimize coupling. According to our results, the variations are statistically significant and the magnitude of the differences is large for both metrics.  NOI, which represents the number of outgoing invocations, has also decreased, and Cliff’s delta value
indicates a small effect size. However, NII exhibits the opposite variation, and the effect size is large.

The manual inspection of the refactored code indicates that developers typically decrease coupling by reducing (1) the strength of dependencies that exist between classes, (2) the message flow of the classes, and (3) the number of inputs a method uses plus the number of subprograms that call this method. The code was improved as expected from the developer's intentions in their commit message.

\noindent\textbf{Example (Positive Impact):} One of the examples showing an improvement in coupling was found in the Maven project\footnote{\textcolor{black}{\url{https://github.com/apache/maven-surefire/commit/d5de47a4f790ea2d18edb5e05c1ef2adcd2db8a2}}}. Developers applied `Move Class' refactoring to move \texttt{JUnit4TestCheckerTest.MySuite2} to \texttt{JUnit3Test\break CheckerTest.MySuite2}. This results in its CBO dropping from 1 to 0, and its RFC from 2 to 1.  The improvement in the CBO and RFC metrics after the removal of code duplication can be related to the elimination of external dependencies and the simplification of method interactions. However, its NII and NOI remain unchanged.

\noindent\textbf{Example (Negative Impact):} One of the examples showing an increase in coupling was found in the Archiva project\footnote{\textcolor{black}{\url{https://github.com/apache/archiva/commit/26e9c3b257bed850d0e2f0bc9dc2d7f11381b789}}}. The developer applied `Extract Superclass' refactoring to extract \texttt{AbstractDiscoverer} from \texttt{AbstractArtifact\break Discoverer}. This results in its CBO increasing from 0 to 1, its RFC from 4 to 6, and its NOI from 0 to 2. However, its NII improves from 3 to 0. The lack of improvement in CBO, RFC, and NOI metrics after the removal of code duplication could be attributed to the specific nature of the duplication and its limited impact on class interactions, method hierarchies, and message handling. 
\noindent\textbf{\textcolor{black}{Complexity.}} Regarding complexity metrics, we consider 16 literature metrics, shown in Table \ref{Table:Quality Metrics Used in This Study.}, to investigate the removal of duplicate code as perceived by developers. As seen in the boxplots in Figures \ref{BP:Complexity-cc}, \ref{BP:Complexity-wmc}, \ref{BP:Complexity-nl}, \ref{BP:coupling-nle}, \ref{BP:Complexity-hcpl}, \ref{BP:Complexity-hdif}, \ref{BP:Complexityheff}, \ref{BP:Complexity-hndb}, \ref{BP:Complexity-hpl}, \ref{BP:Complexity-hpv}, \ref{BP:Complexity-htrp}, \ref{BP:Complexity-hvol}, \ref{BP:Complexity-mims}, \ref{BP:Complexity-mi}, \ref{BP:Complexity-misei}, and \ref{BP:Complexity-mism},  we observe that CC, NL, and NLE remain unchanged, whereas the other 13 metrics experienced an improvement in the median values. The refactored duplicate code exhibits higher values for the four maintenance index-related complexity (\ie MIMS, MI, MISEI, and MISM). The higher values are desirable for these metrics, as shown in Table \ref{Table:Quality Metrics Used in This Study.}. Additionally, the duplicate code refactored shows lower values for the other metrics (\ie WMC, HCPL, HDIF, HEEF, HNDB, HPL, HPV, HTRP, and HVOL), where lower values are desirable after the application of refactoring.



In particular, through a manual inspection of the collected dataset, we observe that developers tend to reduce the number of local methods, simplify the structure statements, reduce the number of paths in the body of the code, and lower the nesting level of the control statements (\eg selection and loop statements) in the method body. 

As seen in Table \ref{Table:Metrics Suites and Metrics Tools Summary}, the \textit{p}-values obtained from all complexity metrics are statistically significant. The effect sizes
calculated in Cliff 's delta ($\delta$) are found to be large only for WMC, small for CC, MISEI, and MISM, and negligible for the remaining 12 metrics.

\noindent\textbf{Example (Positive Impact):} As an illustrative example, we refer to commit\footnote{\textcolor{black}{\url{https://github.com/apache/maven-surefire/commit/d5de47a4f790ea2d18edb5e05c1ef2adcd2db8a2}}} which implements `Move Class' refactoring to move \texttt{JUnit4TestChecker\break Test.MySuite2} to \texttt{JUnit3TestCheckerTest.MySuite2}. Its CC, NL, and NLE remain unaffected, and its WMC improves from 2 to 1. The unchanged CC could be due to the specific nature of the duplicated code, which might not have affected the control flow patterns significantly. However, the improved WMC could be due to consolidation, optimization, or simplification of methods due to the removal of duplicates. In another example\footnote{\textcolor{black}{\url{https://github.com/apache/kafka/commit/f7b7b4745541a576eb0219468263487b07bac959}}}, `Extract Method' refactoring has been applied by developers to extract \texttt{resume} from \texttt{addStreamTasks} to eliminate duplication. Its four maintainability index metrics, \ie MIMS, MI, MISEI, and MISM improved (44.59 to 75.78, 76.25 to 129.6, 56.64 to 153.82, and 33.12 to 89.95), respectively. The remaining complexity metrics, \ie HCPL, HDIF, HNDB, HPL, HPV, HTRP, HVOL, have also improved (343.36 to 23.50, 52.85 to 3, 1515.44 to 44, 184 to 11, 67 to 10, 3277.43 to 6.09, and 1116.16 to 36.54, respectively).

\noindent\textbf{Example (Negative Impact):} As an illustrative example, we refer to commit\footnote{\textcolor{black}{\url{https://github.com/apache/sis/commit/c8ffc0116b86f39caa3d2f45dca5dec68049c93e}}} which implements `Extract Superclass' refactoring to extract \texttt{Element} from \texttt{Copyright} and \texttt{Person}. Its CC increases from 0 to 0.11, its WMC increases from 3 to 45, its NL and NLE increase from 0 to 2. When referring to commit\footnote{\textcolor{black}{\url{https://github.com/apache/ant-ivy/commit/b74264847ef8e9ffeaf06d5fa1fdead4a065b480}}}, the `Extract Method' refactoring to extract \texttt{getReportFile} from \texttt{getRealDependencyRevisionIds} to remove duplication. Its four maintainability index metrics have not improved (75.71 to 64.27, 129.47 to 109.90, 155.64 to 83.06, 91,02, for MIMS, MI, MISEI, and MISM, respectively).   The remaining complexity metrics, \ie HCPL, HDIF, HNDB, HPL, HPV, HTRP, HVOL, have also not improved (61.30 to 120.76, 12.25 to 21, 108.09 to 317.87, 22 to 55, 18 to 30, 62.43 to 314.85, and 91.73 to 269.87, respectively). The absence of improvement in complexity metrics could be due to factors such as the nature of duplicated code, the distribution of complexity across the codebase, and the potential compensatory complexity introduced during the code duplicate removal process.


\noindent\textbf{\textcolor{black}{Inheritance.}} For commits that involve the removal of code duplication, the boxplots depicted in Figures  \ref{BP:Inheritance-dit}, \ref{BP:Inheritance-noc} and \ref{BP:Inheritance-noa} showcase the pre- and post-refactoring results of three structural metrics:  \ie DIT, NOC, and NOA, used in the literature to estimate the inheritance. We observe that only one metric among the three experienced a degradation in median values. Specifically, the median for NOC decreased from 3 to 0, while the median for DIT and NOA increased from 2 to 3 and from 3 to 4, respectively. This suggests that developers may be increasing the depth of the hierarchy by adding more methods for a class to inherit, reducing the number of immediate subclasses, and increasing the number of methods added by a subclass. While some instances show improvement in inheritance, the overall depth of the inheritance tree and the number of methods added by a subclass did not decrease.  The interpretation of the metric improvement depends highly on the quality of the code and the developer's design decisions. The statistical test shows that the differences are statistically significant for DIT, NOC, and NOA. The magnitude of the difference between the three metrics is large.

\noindent\textbf{Example (Positive Impact):} One of the examples that demonstrated improvement in inheritance was found
in a particular commit in the Maven project\footnote{\textcolor{black}{\url{https://github.com/apache/maven-surefire/commit/d5de47a4f790ea2d18edb5e05c1ef2adcd2db8a2}}}.  The developer applied `Move Class' refactoring to move \texttt{JUnit4TestChecker\break Test.CustomSuiteOnlyTest} to \texttt{JUnit3TestCheckerTest.CustomSuiteOnlyTest}. Its DIT drops from 1 to 0, its NOC remains unaffected, and its NOA improves from 1 to 0. This increases the reuse of common code logic and leads to more effective inheritance relationships and a better-defined hierarchy.

\noindent\textbf{Example (Negative Impact):} One of the examples that showed improvement in inheritance was found
in a particular commit in the Archiva project\footnote{\textcolor{black}{\url{https://github.com/apache/archiva/commit/26e9c3b257bed850d0e2f0bc9dc2d7f11381b789}}}.  The developer applied `Extract Superclass' refactoring to extract \texttt{AbstractDiscoverer} from class \texttt{AbstractArtifactDiscoverer}. Its DIT increases from 0 to 1, its NOC remains unaffected, and its NOA increases from 0 to 1. This indicates that the refactoring applied to remove duplication does not always improve inheritance metrics due to either pre-existing inheritance challenges, or the focused nature of the duplication removal. 

\noindent\textbf{\textcolor{black}{Design Size.}} For commits whose messages report the removal of code duplicate, the boxplots sketched in Figures \ref{BP:Design Size-loc}, \ref{BP:Design Size-tloc}, \ref{BP:Design Size-lloc}, \ref{BP:Design Size-tlloc}, \ref{BP:Design Size-cloc},  \ref{BP:Design Size-npm}, \ref{BP:Design Size-nos}, and \ref{BP:Design Size-tnos}  show the pre- and post-refactoring results of four structural metrics, \ie LOC, TLOC, LLOC, TLLOC, CLOC,  NPM, NOS, and TNOS, used in the literature to estimate the design size. We notice the improvement of six metrics, namely LOC, TLOC, LLOC, TLLOC, NOS, and TNOS after the commits in which developers explicitly target the improvement of code duplication, their variations are statistically significant.  
 The magnitude of LOC, TLOC, and TLLOC is small, whereas the magnitude for LLOC, NOS, and TNOS is negligible. As seen in the box plots, the medians generally decreased. However, we note that the medians for CLOC and NPM remain unchanged. The differences in CLOC and NPM are statistically significant, and the magnitude of the difference is negligible and large, respectively. This indicates that developers generally retain the lines containing comments and maintain the same number of methods after applying refactoring. 

\noindent\textbf{Example (Positive Impact):} As an illustrative example, we refer to the commit\footnote{\textcolor{black}{\url{https://github.com/apache/maven-surefire/commit/d5de47a4f790ea2d18edb5e05c1ef2adcd2db8a2}}} which implements `Extract Method' refactoring to extract \texttt{accept\break(testClass)} from \texttt{invalidTest}. Its LOC, TLOC, LLOC, TLLOC drop from 6 to 4, and its CLOC, NOS, and TNOS remain unaffected.  Furthermore, when moving the class \texttt{JUnit4TestCheckerTest.AlsoValid}  to \texttt{JUnit3TestChecker\break Test.AlsoValid}, its NPM improves from 1 to 0.  In qualitative terms, the removal of code duplication and the introduction of a dedicated method have led to more modular, focused, and readable code. This shows that size metrics capture the removal of code duplication as perceived by the developer.

\noindent\textbf{Example (Negative Impact):} As an illustrative example, we refer to the commit\footnote{\textcolor{black}{\url{https://github.com/apache/commons-bcel/commit/67dfdf60f5f8ccb8ed910bfe9d1cdc6e84f0db36}}} which implements `Extract Method' refactoring to extract \texttt{accept\break(createAnnotationEntries)} from \texttt{getAnnotationEntries}. Its LOC and TLOC increased from 7 to 11, its LLOC, TLLOC increased from 6 to 10, and its NOS and TNOS increased from 3 to 6. Its CLOC decreases from 3 to 1. The observed lack of improvement, in this case, can be attributed to a couple of factors, including the nature of the changes made, the extent of duplication and additional compensatory changes. This results in an overall increase in the class size as assessed by these employed design size metrics.

\noindent\textbf{\textcolor{black}{Summary.}} \textcolor{black}{This \textcolor{black}{section} summarizes our findings and their implications.}

\begin{itemize}
    \item \textcolor{black}{\textbf{Cohesion.} The normalized LCOM metric not only serves as a suitable substitute for the original LCOM but also serves as a representation of the cohesion quality attribute. A positive variation in this metric aligns with the developer's intention to eliminate code duplication.}
    \item \textcolor{black}{\textbf{Coupling.} CBO, RFC, and NOI generally improve as the developer intends to eliminate code duplication, and their variation is significant. NII exhibits opposite variations in
coupling.}
    \item \textcolor{black}{\textbf{Complexity.} CC, NL, and NLE remain unchanged, and the remaining 13 complexity-related metrics generally improve as the developer intends to improve code duplicate, and all their variation is significant.}
    \item \textcolor{black}{\textbf{Inheritance.} NOC generally decreases as the developer intends to remove code duplication, and its variation is significant. DIT and NOA exhibit opposite variations in inheritance.}
    \item \textcolor{black}{\textbf{Design Size.} LOC, TLOC, LLOC, TLLOC, NOS, TNOS generally improve as developers intend to remove code duplication, and their variations are significant. These metrics have a significant positive variation which matches the developer's perception of removing code duplicates.}
\end{itemize}


\subsection{What are the refactoring operations that are associated with code duplicate removal?}
\begin{figure}[t]
\centering 
\begin{tikzpicture}
\begin{scope}[scale=0.8]
\pie[rotate = 180,pos ={0,0},text=inside,outside under=20,no number]{55.7/Extract Method\and55.7\%, 37.5/Move Method\and37.5\%, 3.8/Extract Superclass\and3.8\%,2.7/Move Attribute\and2.7\%,0.3/Move Class\and0.3\%}
\end{scope}
\end{tikzpicture}
\caption{Distribution of refactoring operations for code duplicate removal.}
\label{fig:refactoringtypes}
\end{figure}
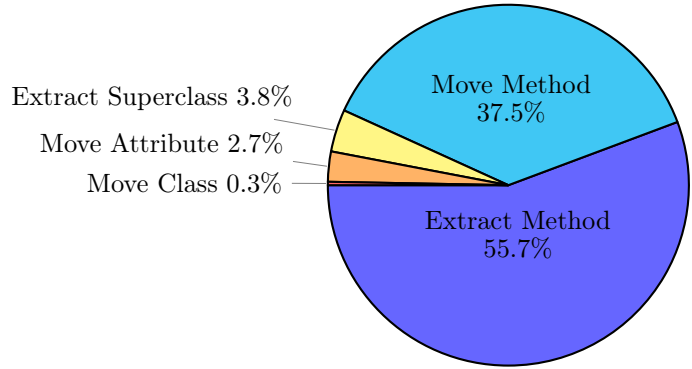
Looking at the refactoring operations that could play a role in code duplicate removal, Figure \ref{fig:refactoringtypes} depicts the percentages of refactoring operations. \textcolor{black}{As can be seen, the most common category concerns `Extract Method', representing 55.7\% of the commits. This observation is in line with the findings of previous studies describing that `Extract Method' refactoring is considered \say{Swiss army knife} of
refactorings as developers often apply it to eliminate duplicated code \citep{higo2004aries,higo2005aries,higo2008metric,tairas2012increasing,bian2013spape,yue2018automatic,yoshida2019proactive,arcelli2015duplicated,alomar2022anticopypaster,alomar2023just,alomar2024behind}. In fact, a recent study on extract method refactoring highlights that method extraction is one of the main refactorings that were defined when the area was established \citep{alomar2024behind,griswold1993automated}, as it is a common response to the need to keep methods concise and modular, and reduced the spread of shared responsibilities.} The next most common categories are `Move Method', representing 37.5\% of the commits. This indicates that developers might improve the quality of the code by moving the method containing duplication to a different class, effectively eliminating duplicated code. The category `Extract Superclass', `Move Attribute', and `Move Class' had the least number of commits, which had a ratio of 3.8\%, 2.7\%, and 0.3\%, respectively. 

When performing manual inspection of source code, we notice that these five refactoring operations contribute to the elimination of code duplication in several ways. By performing the `Extract Method' refactoring, redundant code segments can be consolidated into a single method that can be reused across different parts of the codebase. Additionally, when moving methods from one class to another using `Move Method' refactoring, it helps centralize logic and eliminate duplicate code that might have been present in multiple classes. Moreover, by extracting a superclass using `Extract Superclass' refactoring, it encapsulates common attributes and behaviors of related classes, allowing duplicated code to be consolidated. This can be followed by moving shared attributes to a common superclass using `Move Attribute' refactoring to reduce redundancy and ensures that changes to these attributes are reflected across all subclasses. Finally, moving the entire class to a common location using `Move Class' refactoring can help in reducing duplicated code, and it is useful when classes share similar functionality but exist in different parts of the codebase.


\section{Lessons Learned}
\label{Section:lesson}

\begin{figure*}[htbp]
\centering 
\includegraphics[width=1.2\textwidth]{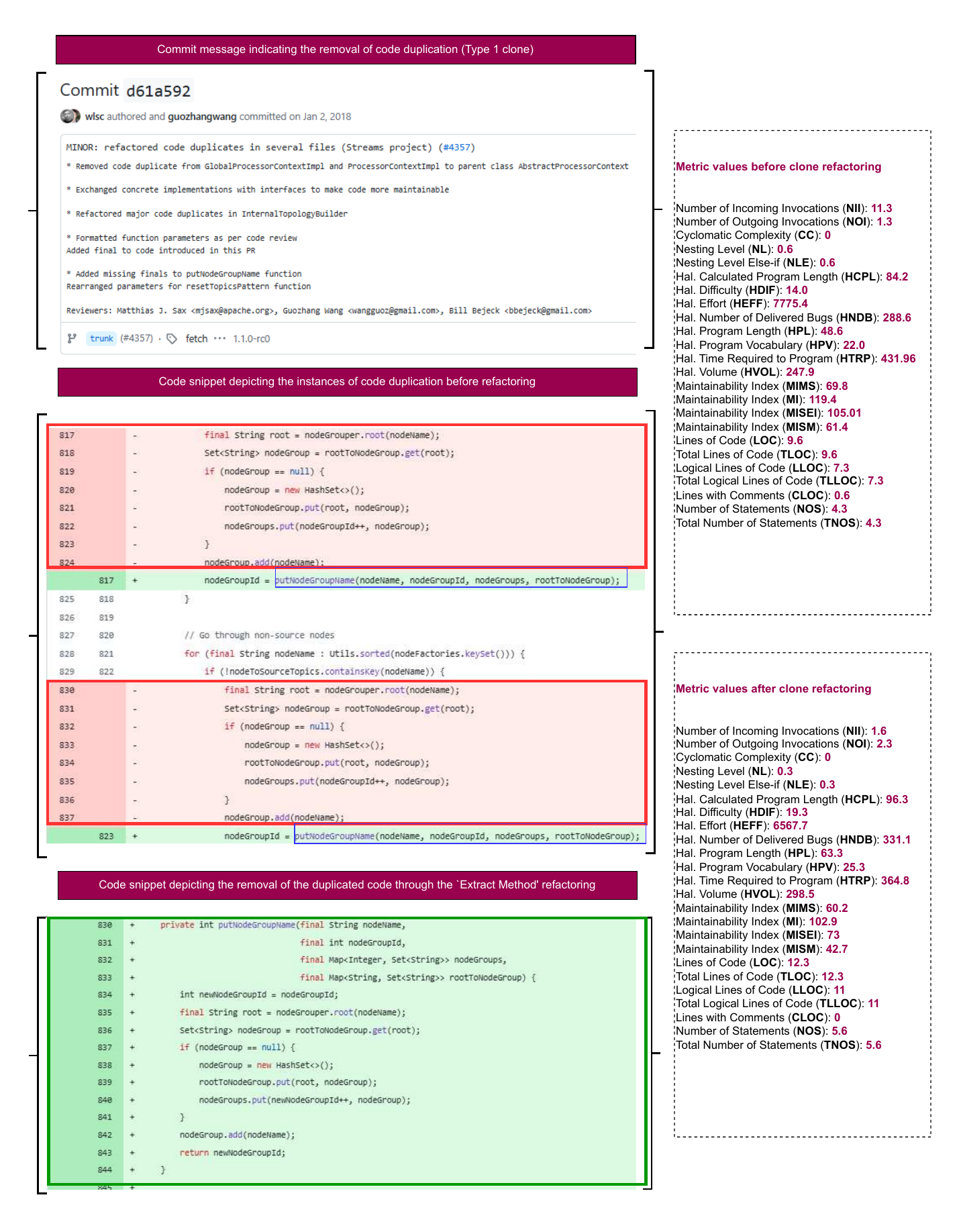}
\caption{\textcolor{black}{Example of selected Type-1 code clone from kafka project.}} 
\label{fig:example-casestudies-type1}
\end{figure*}

 \begin{figure*}[htbp]
\centering 
 \includegraphics[width=1.2\textwidth]{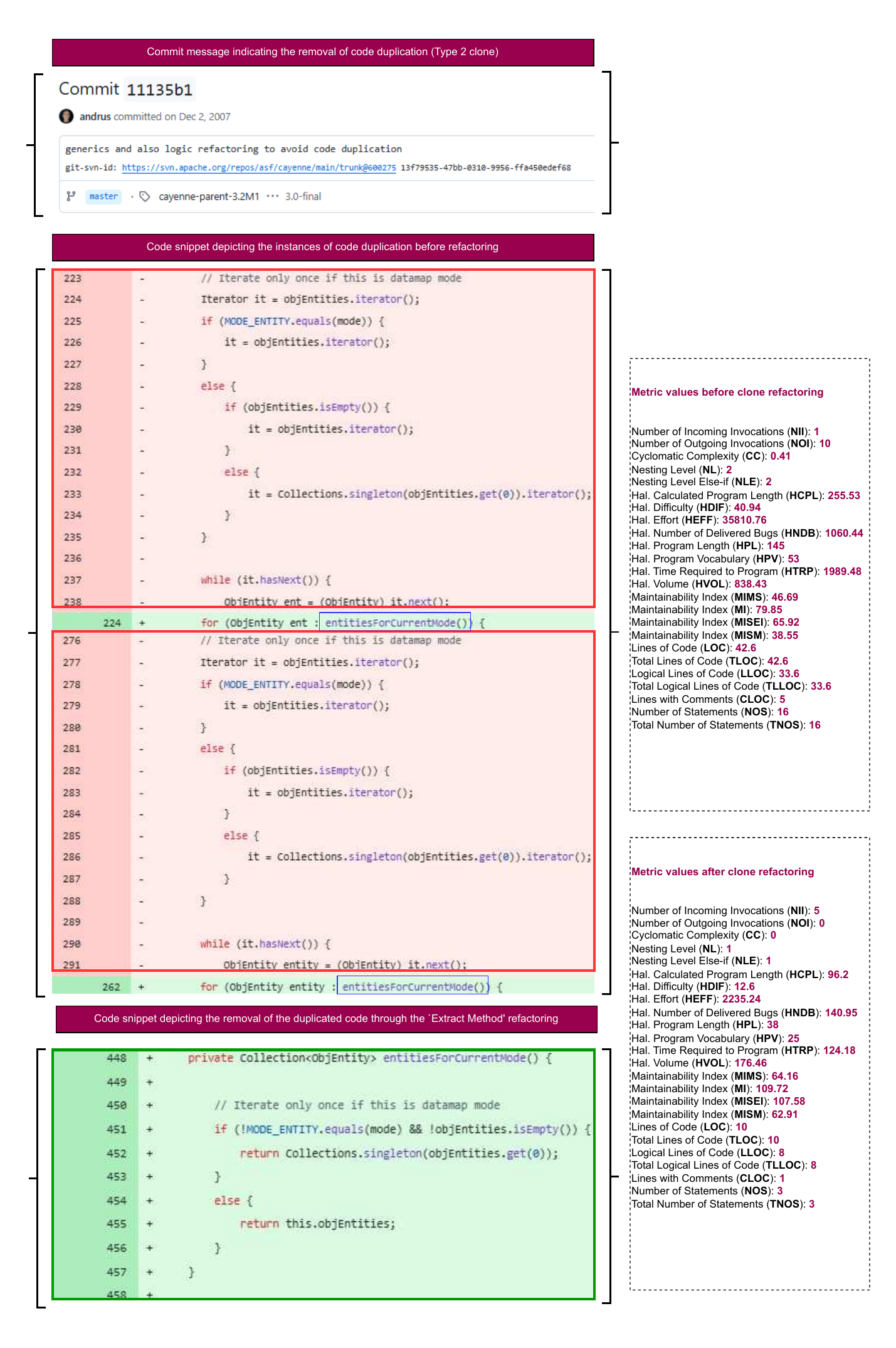}
 \caption{\textcolor{black}{Example of selected Type-2 code clone from cayenne project.}} 
 \label{fig:example-casestudies-type2}
\end{figure*}

\begin{figure*}[htbp]
\centering 
\includegraphics[width=1.3\textwidth]{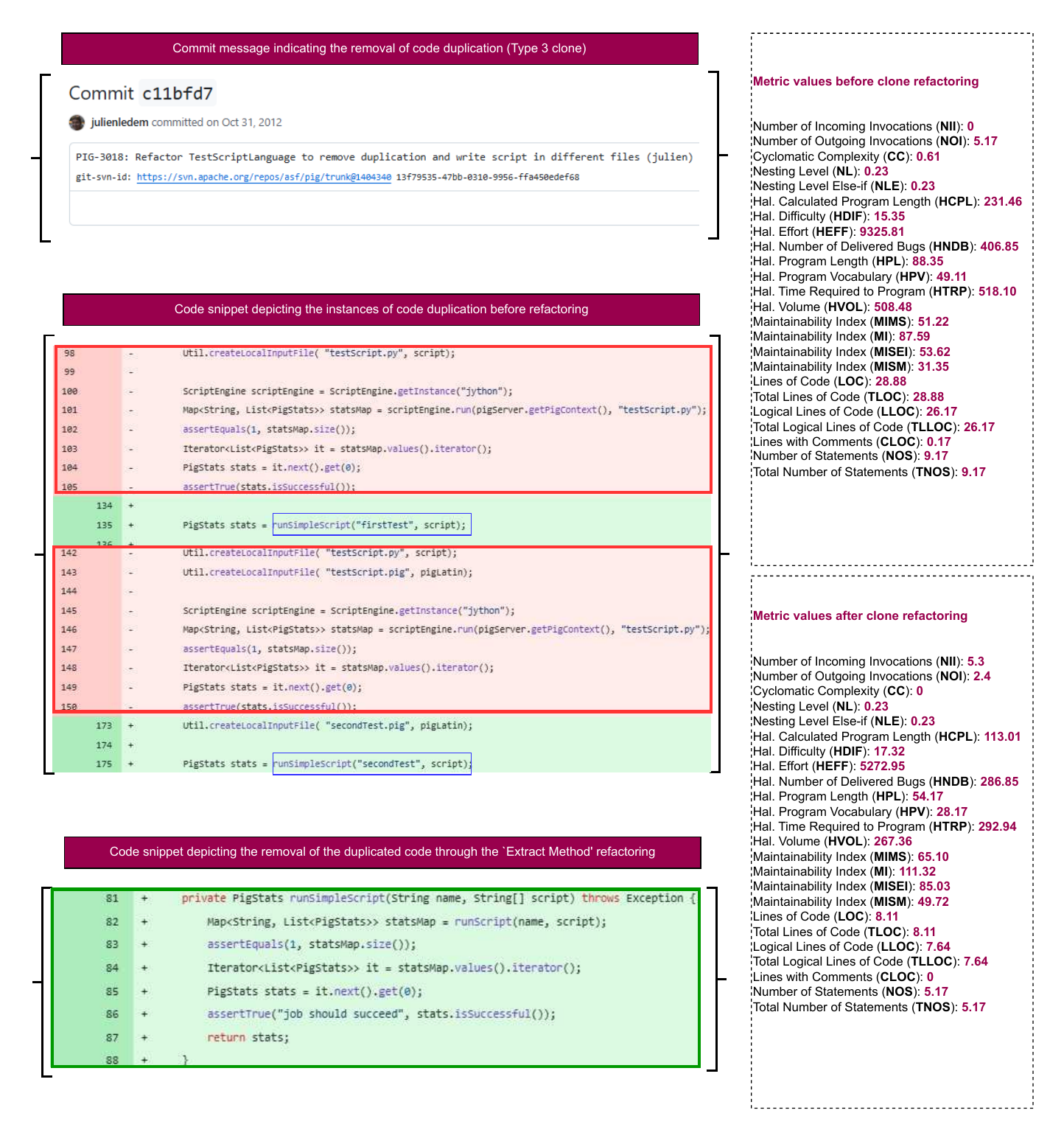}
\caption{\textcolor{black}{Example of selected Type-3 code clone from pig project.}} 
\label{fig:example-casestudies-type3}
\end{figure*}

\noindent{\textbf{ \textcolor{black}{Lesson 1: Code clones associated with commits about duplicate removal are from different clone types.}}}  \textcolor{black}{There are various types of code clone exist in the literature (\ie Type-1, Type-2, Type-3, and Type-4) \citep{mondal2020survey}. When performing manual examination of commits associated with code clones, we realized that some commits with the explicit intention of removing duplication are associated with different clone types. Furthermore, in some commits associated with duplicate removal, developers can combine clone refactoring with other unrelated changes, such as feature updates, bug fixes, or general code cleanup. This observation is consistent with existing studies that show that developers interleave refactoring with other changes, and 11– 39\% of bug fixing commits include other changes \cite{silva2016we,alomar2021we,murphy2012we,nguyen2013filtering}.}


\noindent{\textbf{ \textcolor{black}{Lesson 2: Refactoring different types of clones can have different variations on metric values.}}}  \textcolor{black}{As illustrated in \textcolor{black}{Tables \ref{Table:Quality Metrics in Related Work} and \ref{Table:Quality Metrics in Related Work-v2}}, there have been two decades' worth of work on the relationship between refactoring and code quality. We can see that there is room for empirical investigation of the impact of clone removal refactorings on internal quality metrics. In this study, we observe that the impact of refactoring clones on software quality metrics can vary based on the type of clone being refactored. Moreover, developers may have various mechanisms that contribute to removing duplicates, and these strategies may dictate different variations on the metrics. However, locating refactored clone types for each instance presents multiple challenges: (1) a single commit can address multiple clone types simultaneously, making it difficult to attribute metric variations to a specific clone type;  (2) some clone types may occur less frequently in the dataset, further complicating efforts to draw conclusions regarding the influence of clone types on metric variations; and (3) manually determining clone types for each instance is time-consuming and prone to error, particularly when dealing with a large dataset. Although existing clone detection tools can detect the clone, they require additional configuration and setup by the users. In the following, we show an example of each type of clone and its refactoring:}
\begin{itemize}
    \item \textcolor{black}{\textit{Type-1 code clone.} Figure \ref{fig:example-casestudies-type1} illustrates a Type-1 clone that has been refactored. The example demonstrates two duplicate instances, which represent a Type-1 clone (\ie identical code fragments). An `Extract Method' refactoring was applied, resulting in the extraction of the method \texttt{putNodeGroup\break Name(nodeName String, nodeGroupId int, nodeGroups Map, rootTo\break NodeGroup Map)} from \texttt{makeNodeGroups()} in the \texttt{InternalTopologyBuilder} class. For the complexity metrics, we observed varied behavior: CC remained unchanged, some metrics showed improvement (NL, NLE, HEFF, HPL, and HTRP), while others did not improve (HCPL, HDIF, HNDB, HPV, HVOL, MIMS, MI, MISEI, and MISM). Regarding the size metrics, none showed improvement. For coupling metrics, NII improved, whereas NOI did not.} 
    \item \textcolor{black}{\textit{Type-2 code clone.} Figure \ref{fig:example-casestudies-type2} depicts a Type-2 clone that has been refactored. This example highlights two duplicate instances, categorized as a Type-2 clone (\ie syntactically identical fragments). The method \texttt{entitiesForCurrentMode()} was extracted from \texttt{generateClassPairs\_1\_1\break (classTemplate String, superTemplate String, superPrefix String)} in the \texttt{MapClassGenerator} class using the `Extract Method' refactoring operation. The complexity metrics have shown improvement, while the design size metrics have also improved, with the exception of CLOC. For coupling metrics, NOI has improved, whereas NII has not.}
    \item \textcolor{black}{\textit{Type-3 code clone.} Figure \ref{fig:example-casestudies-type3} shows a Type-3 clone that has been refactored. The example illustrates two duplicate instances, identified as a Type-3 clone (\ie copied fragments with further modifications such as changed, added, or removed statements). Through the `Extract Method' refactoring, the method \texttt{runSimpleScript(String name, String[] script)}  was extracted in the \texttt{TestScriptLanguage} class. The complexity metrics have improved overall, with the exception of HDIF, which has decreased, while NL and NLE remain unchanged. Size metrics have also improved, except for CLOC. For coupling metrics, NOI has improved, but NII has decreased. }
\end{itemize}

\begin{figure*}[htbp]
\centering 
\includegraphics[width=1.2\textwidth]{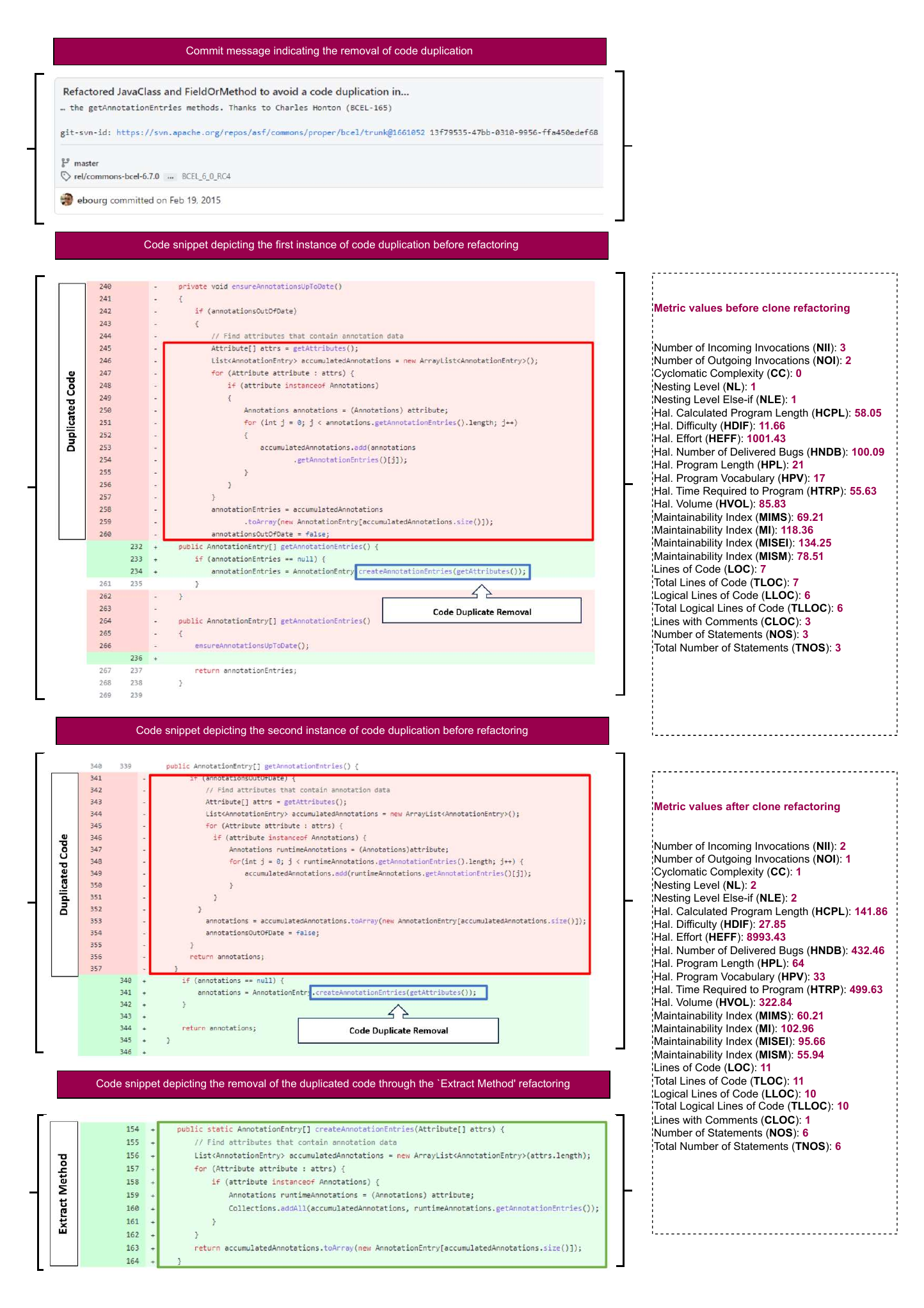}

 \caption{\textcolor{black}{Example of selected commit message from commons-bcel project.}}
 \label{fig:example-casestudies-2}
\end{figure*}

\noindent{\textbf{ \textcolor{black}{Lesson 3: Some state-of-the-art metrics can capture the developer’s intention of removing code duplication with different degrees of improvement and degradation of software quality.}}} \textcolor{black}{When removing code duplication, developers often perform `Extract Method' refactoring with the expectation of improving code quality. Yet, the state-of-the-art metrics may reflect varying levels of improvement or even degradation following these refactoring events.  For example, in Figure \ref{fig:example-casestudies-2}, we demonstrate the code snippet depicting the instances of code duplication before and after refactoring. We can see that refactoring mining tools detect `Extract Method' refactoring from project commoms-bcel\footnote{\textcolor{black}{\url{https://github.com/apache/commons-bcel/commit/67dfdf60f5f8ccb8ed910bfe9d1cdc6e84f0db36}}}  to extract \texttt{createAnnotationEntries} from \texttt{getAnnotationEntries}. This example emphasizes how refactoring can have mixed effects, positively influencing some metrics while negatively impacting others. As can be seen, its coupling metrics (NII and NOI) have been improved. However, its complexity metrics (CC, NL, NLE, HCPL, HDIF, HEFF, HNDB, HPL, HPV, HTRP, HVOL, MIMS, MI, MISEI, and MISM) and size metrics (LOC, TLOC, LLOC, TLLOC, CLOC, NOS, and TNOS) have not been improved. For metrics where the metrics do not capture the developer's intention, several possible explanations can be consideblack: }
\begin{itemize}
    \item \textcolor{black}{\textit{Inadequacy of the metrics for certain scenarios.} The metrics used to assess software quality, may not always be the most suitable for reflecting the specific intention behind a refactoring. For instance, a developer may intend to improve readability or maintainability, but standard structural metrics may not effectively quantify these aspects. This misalignment between developer goals and the measublack outcomes can lead to discrepancies in how the impact of refactoring is perceived.}
    \item \textcolor{black}{\textit{Limitations of the metrics.} The state-of-the-art metrics have inherent limitations and may not comprehensively capture the effects of refactoring. For example, metrics such as CC focus on the control flow but may overlook improvements in code modularity. This indicates a need to either refine existing metrics or introduce new ones that better align with developer goals, particularly in cases of complex refactoring.}
    \item \textcolor{black}{\textit{Deviation from developer intentions.} In some cases, developers' intentions, as stated in commit messages, may not align with the actual changes performed in the codebase. This could happen for various reasons. For example, a commit message may report the removal of duplicate code, but the implementation might only partially address the duplication or introduce new dependencies, resulting in no measurable improvement or even metric degradation.}
\end{itemize}

\section{Implications}
\label{Section:Implication}


 \noindent{\textbf{ Further advancing quality metrics and duplicate code removal.}} The existing literature discusses various automatic refactoring approaches aimed at assisting practitioners in detecting antipatterns or code smells. Baqais and Alshayeb \citep{baqais2020automatic} have highlighted the growing interest in automatic refactoring studies. The researchers explored the potential of machine learning to identify refactoring opportunities. Since features play a vital role in the quality of machine learning models obtained, this study can contribute to determining which metrics can serve as effective features in machine learning algorithms, facilitating the accurate recommendation of refactoring opportunities at different levels of granularity (\ie class, method, field), which can assist developers in automatically making their decisions. For example, incorporating the most impactful metrics as features in predicting whether a given piece of code should undergo a specific refactoring operation can enhance developers' confidence in accepting recommended refactorings or selecting the most suitable refactoring candidate. This knowledge is needed because, in practice, the built model should require as little data as possible. Furthermore, since we observe that some of the quality metrics did not capture any improvement, we plan to conduct more experiments to validate the effectiveness of these metrics to explore whether the observations are due to the appropriateness of the quality metrics or to the needed validation and clarity of the perception of the developers.

 \noindent{\textbf{ Putting developer in the loop when designing refactoring recommendation systems.}} Based on the findings, it becomes evident that different structural metrics have the capacity to depict code duplication, thereby influencing software quality in diverse ways. Certain metrics improve software quality, whereas others might result in its decline. 
 This underscores the importance of involving developers in the design of refactoring recommendation systems, effectively engaging them in the process. This approach emerges as effective in discerning meaningful refactorings that align with the perspectives of developers \citep{hall2012supervised,bavota2012putting,pantiuchina2018improving}.

 \noindent{\textbf{ Examining the code duplicate removal potentials with refactoring.} Our study reveals the context in which developers refactor the code to eliminate code duplicates. Our future research direction can focus on providing a comprehensive taxonomy for code duplication-aware refactoring practices. This taxonomy can show various contexts of code duplicates and refactoring and can demonstrate different forms of code reuse. Thereafter, researchers can build on top of our findings to better understand developer practices and investigate to what extent this taxonomy for refactoring with awareness of duplicate code improves the system's quality.

 \noindent\textbf{ Understanding the completeness of the quality metric capturing duplicate code removal as documented by developers.} We observe that not all quality metrics can capture the improvement in duplicate code removal perceived by developers in their commit messages. Although quality metrics can help pinpoint design flaws for refactoring recommendation systems, such a recommendation would be meaningful if qualitative insights from developers complemented it. Furthermore, the alignment or disparity between the enhancement of software quality as perceived by developers and its evaluation through quality metrics can be attributed to factors such as the focused nature of the duplication removal, the extent of duplication, and the potential compensatory changes. Future research is encouraged to consider the direct effect of duplicate removal and the broader context of code changes and their implications for quality metrics. 


 \noindent{\textbf{ Investigating the characteristics and effects of eliminating code duplication on software quality.} The results advance our understanding of the effects of eliminating code duplication on software quality. It is evident that certain software quality metrics can be used as indicators for code fragments that are more likely to be extracted and identified as problematic and should be removed by refactoring. Consequently, a threshold can be established to show when quality metrics reach a level where duplicate code will have a negative effect on maintenance and need to be refactored.

\section{Threats to Validity}
\label{Section:Threats}

In this section, we describe potential threats to the validity of our research method and the actions we took to mitigate them.

\textbf{Internal Validity.} The accuracy of our analysis is primarily dependent on the precision of the refactoring mining tools, as these tools may miss the detection of some refactorings. However, previous studies \citep{silva2016we,tsantalis2018accurate,silva2017refdiff} report that \texttt{RefactoringMiner} and \texttt{RefDiff} have high precision and recall scores compared to other state-of-the-art refactoring detection tools, giving us confidence in using the tools. Another potential threat to validity is related to commit messages. \textcolor{black}{This study does not exclude commits containing tangle code changes \citep{herzig2016impact,kirinuki2014hey}, where developers made changes related to different tasks and one of these tasks could be related to quality improvement. If these changes were committed at once, there is a possibility that the individual changes merge and that the original task cannot be traced back. Similarly to the previous study \cite{pantiuchina2018improving}, we did not consider filtering out such changes in this study}. Moreover, our manual analysis is time-consuming and error-prone, which we tried to mitigate by focusing mainly on commits known to contain refactorings. 

Another potential threat to validity is sample bias, where the choice of the data can directly impact the results. Therefore, we explored a large sample of projects from the SmartSHARK dataset \citep{trautsch2021msr}, to ensure the quality of the findings and diversify the sources to reduce the bias of the data belonging to the same entity. The qualitative analysis was conducted by a single author, which could introduce bias into the process. However, commits that were debatable were discarded. We also provide our dataset online for further refinement and analysis. 

\textbf{Construct Validity.} A potential threat to construct validity relates to the set of metrics, as it may miss some properties of the selected internal quality attributes. To address this potential threat, we mitigate it by choosing well-known metrics that encompass various properties of each attribute, as reported in the literature \citep{chidamber1994metrics}.

\textbf{External Validity.} Our analysis was limited to only open-source Java projects. However, we were able to examine 128 projects, which were well-commented and exhibited diversity in terms of size, contributors, number of commits, and refactorings. \textcolor{black}{Still, we believe that the results found in this study are largely language-agnostic. However, certain language-specific characteristics, such as syntax complexity and tooling support, can influence duplication patterns. Although we expect similar trends across languages with similar paradigms, a comprehensive analysis encompassing various languages is recommended to confirm this generalization.}

\section{Conclusion}
\label{Section:Conclusion}

We conducted an empirical study to investigate the alignment between code duplicate removal and software design metrics focusing on 5 internal quality attributes and 32 structural metrics. In particular, we obtained a corpus of more than two million refactoring commits from 128 open-source Java projects. We then extracted 32 structural metrics to identify code duplicate removal commits and the refactoring operations associated with them. \textcolor{black}{ In summary, the main conclusions are:}

\textemdash \textcolor{black}{Our findings show that some state-of-the-art metrics can capture the developer's intention of removing code duplication with different degrees of improvement
and degradation of software quality.}

\textemdash \textcolor{black}{Many metrics associated with key quality attributes, such as cohesion, coupling, complexity, and size, reflect the developers' intentions for duplicate removal mentioned in commit messages. In contrast, there are instances where the metrics do not represent the quality improvements stated by the developers.}

\textemdash \textcolor{black}{As for inheritance, NOC generally decreases as the developer intends to remove code duplication. DIT and NOA exhibit
opposite variations in inheritance, so these findings motivate a deeper investigation to understand the mismatch between theory and practice.}

\textcolor{black}{\textit{Implications.} As most of the mapped metrics associated with the main quality attributes successfully capture developers’ intentions for removing code duplicates, as is evident from the commit messages, we believe our study enables the following novel applications:}

\textemdash \textcolor{black}{Given that features significantly influence the quality of machine learning models, we can help identify which metrics may function as effective features within these algorithms, thus supporting developers in their decision-making. Using the most influential metrics as features to predict whether a particular code fragment should be subject to a specific refactoring operation, we can enhance developers’ confidence in accepting recommended refactorings or selecting the most suitable refactoring candidate.}

\textemdash \textcolor{black}{Empirical researchers can focus on providing a comprehensive taxonomy for code duplication-aware refactoring practices, showing various contexts of code duplicates and refactoring and demonstrating different forms of code reuse.}

\textemdash \textcolor{black}{A qualitative investigation through a developer survey can be conducted to better understand the motivation behind refactoring activities in the context of code duplicate removal to improve software metrics.}



\section{Acknowledgments}
\noindent\textbf{Declaration of generative AI and AI-assisted technologies in the writing process.}
During the preparation of this work, the author used the ChatGPT web interface and the Wrietfull tool to improve the language and readability of the manuscript. After using this tool, the author reviewed and edited the content as needed and takes full responsibility for the content of the publication.

{\footnotesize\bibliography{references.bib}}

\begin{thebibliography}{100}
\expandafter\ifx\csname url\endcsname\relax
  \def\url#1{\texttt{#1}}\fi
\expandafter\ifx\csname urlprefix\endcsname\relax\def\urlprefix{URL }\fi
\expandafter\ifx\csname href\endcsname\relax
  \def\href#1#2{#2} \def\path#1{#1}\fi

\bibitem{roy2009comparison}
C.~K. Roy, J.~R. Cordy, R.~Koschke, Comparison and evaluation of code clone detection techniques and tools: A qualitative approach, Science of computer programming 74~(7) (2009) 470--495.

\bibitem{thongtanunam2019will}
P.~Thongtanunam, W.~Shang, A.~E. Hassan, {Will this clone be short-lived? Towards a better understanding of the characteristics of short-lived clones}, Empirical Software Engineering 24~(2) (2019) 937--972.

\bibitem{fanta1999removing}
R.~Fanta, V.~Rajlich, Removing clones from the code, Journal of Software Maintenance: Research and Practice 11~(4) (1999) 223--243.

\bibitem{Fowler:1999:RID:311424}
M.~Fowler, K.~Beck, J.~Brant, W.~Opdyke, d.~Roberts, \href{http://dl.acm.org/citation.cfm?id=311424}{Refactoring: Improving the Design of Existing Code}, Addison-Wesley Longman Publishing Co., Inc., Boston, MA, USA, 1999.
\newline\urlprefix\url{http://dl.acm.org/citation.cfm?id=311424}

\bibitem{silva2016we}
D.~Silva, N.~Tsantalis, M.~T. Valente, Why we refactor? confessions of github contributors, in: Proceedings of the 2016 24th ACM SIGSOFT International Symposium on Foundations of Software Engineering, ACM, 2016, pp. 858--870.

\bibitem{murphy2008breaking}
E.~Murphy-Hill, A.~P. Black, {Breaking the barriers to successful refactoring: Observations and tools for Extract Method}, in: Proceedings of the 30th international conference on Software engineering, 2008, pp. 421--430.

\bibitem{pantiuchina2018improving}
J.~Pantiuchina, M.~Lanza, G.~Bavota, Improving code: The (mis) perception of quality metrics, in: 2018 IEEE International Conference on Software Maintenance and Evolution (ICSME), IEEE, 2018, pp. 80--91.

\bibitem{alomar2019impact}
E.~A. AlOmar, M.~W. Mkaouer, A.~Ouni, M.~Kessentini, On the impact of refactoring on the relationship between quality attributes and design metrics, in: 2019 ACM/IEEE International Symposium on Empirical Software Engineering and Measurement (ESEM), IEEE, 2019, pp. 1--11.

\bibitem{trautsch2021msr}
A.~Trautsch, F.~Trautsch, S.~Herbold, Msr mining challenge: The smartshark repository mining data, arXiv preprint arXiv:2102.11540.

\bibitem{sahraoui2000can}
H.~A. Sahraoui, R.~Godin, T.~Miceli, Can metrics help to bridge the gap between the improvement of oo design quality and its automation?, in: icsm, IEEE, 2000, p. 154.

\bibitem{stroulia2001metrics}
E.~Stroulia, R.~Kapoor, Metrics of refactoring-based development: An experience report, in: OOIS 2001, Springer, 2001, pp. 113--122.

\bibitem{kataoka2002quantitative}
Y.~Kataoka, T.~Imai, H.~Andou, T.~Fukaya, A quantitative evaluation of maintainability enhancement by refactoring, in: International Conference on Software Maintenance, 2002. Proceedings., IEEE, 2002, pp. 576--585.

\bibitem{demeyer2002maintainability}
S.~Demeyer, Maintainability versus performance: What’s the effect of introducing polymorphism, Edegem, Belgium: Universiteit Antwerpe.

\bibitem{tahvildari2003quality}
L.~Tahvildari, K.~Kontogiannis, J.~Mylopoulos, Quality-driven software re-engineering, Journal of Systems and Software 66~(3) (2003) 225--239.

\bibitem{leitch2003assessing}
R.~Leitch, E.~Stroulia, Assessing the maintainability benefits of design restructuring using dependency analysis, in: Proceedings. 5th International Workshop on Enterprise Networking and Computing in Healthcare Industry (IEEE Cat. No. 03EX717), IEEE, 2003, pp. 309--322.

\bibitem{du2003describing}
B.~Du~Bois, T.~Mens, Describing the impact of refactoring on internal program quality, in: International Workshop on Evolution of Large-scale Industrial Software Applications, 2003, pp. 37--48.

\bibitem{tahvildari2003metric}
L.~Tahvildari, K.~Kontogiannis, A metric-based approach to enhance design quality through meta-pattern transformations, in: Seventh European Conference onSoftware Maintenance and Reengineering, 2003. Proceedings., IEEE, 2003, pp. 183--192.

\bibitem{du2004refactoring}
B.~Du~Bois, S.~Demeyer, J.~Verelst, Refactoring-improving coupling and cohesion of existing code, in: 11th working conference on reverse engineering, IEEE, 2004, pp. 144--151.

\bibitem{du2005does}
B.~Du~Bois, S.~Demeyer, J.~Verelst, Does the" refactor to understand" reverse engineering pattern improve program comprehension?, in: Ninth European Conference on Software Maintenance and Reengineering, IEEE, 2005, pp. 334--343.

\bibitem{geppert2005refactoring}
B.~Geppert, A.~Mockus, F.~Robler, Refactoring for changeability: A way to go?, in: 11th IEEE International Software Metrics Symposium (METRICS'05), IEEE, 2005, pp. 10--pp.

\bibitem{ratzinger2005improving}
J.~Ratzinger, M.~Fischer, H.~Gall, Improving evolvability through refactoring, in: Proceedings of the 2005 international workshop on Mining software repositories, 2005, pp. 1--5.

\bibitem{moser2006does}
R.~Moser, A.~Sillitti, P.~Abrahamsson, G.~Succi, Does refactoring improve reusability?, in: International Conference on Software Reuse, Springer, 2006, pp. 287--297.

\bibitem{wilking2007empirical}
D.~Wilking, U.~F. Kahn, S.~Kowalewski, An empirical evaluation of refactoring., e-Informatica 1~(1) (2007) 27--42.

\bibitem{stroggylos2007refactoring}
K.~Stroggylos, D.~Spinellis, Refactoring--does it improve software quality?, in: Fifth International Workshop on Software Quality (WoSQ'07: ICSE Workshops 2007), IEEE, 2007, pp. 10--10.

\bibitem{moser2007case}
R.~Moser, P.~Abrahamsson, W.~Pedrycz, A.~Sillitti, G.~Succi, A case study on the impact of refactoring on quality and productivity in an agile team, in: IFIP Central and East European Conference on Software Engineering Techniques, Springer, 2007, pp. 252--266.

\bibitem{shrivastava2008impact}
S.~V. Shrivastava, V.~Shrivastava, Impact of metrics based refactoring on the software quality: a case study, in: TENCON 2008-2008 IEEE Region 10 Conference, IEEE, 2008, pp. 1--6.

\bibitem{higo2008refactoring}
Y.~Higo, Y.~Matsumoto, S.~Kusumoto, K.~Inoue, Refactoring effect estimation based on complexity metrics, in: 19th Australian Conference on Software Engineering (aswec 2008), IEEE, 2008, pp. 219--228.

\bibitem{Reddy2009quantitative}
K.~N. Reddy, A.~A. Rao, A quantitative evaluation of software quality enhancement by refactoring using dependency oriented complexity metrics, in: 2009 Second International Conference on Emerging Trends in Engineering \& Technology, IEEE, 2009, pp. 1011--1018.

\bibitem{alshayeb2009empirical}
M.~Alshayeb, Empirical investigation of refactoring effect on software quality, Information and software technology 51~(9) (2009) 1319--1326.

\bibitem{alshayeb2009refactoring}
M.~Alshayeb, Refactoring effect on cohesion metrics, in: 2009 International Conference on Computing, Engineering and Information, IEEE, 2009, pp. 3--7.

\bibitem{usha2009quantitative}
K.~Usha, N.~Poonguzhali, E.~Kavitha, A quantitative approach for evaluating the effectiveness of refactoring in software development process, in: 2009 Proceeding of International Conference on Methods and Models in Computer Science (ICM2CS), IEEE, 2009, pp. 1--7.

\bibitem{hegedHus2010effect}
G.~Heged{\H{u}}s, G.~Hrabovszki, D.~Heged{\H{u}}s, I.~Siket, Effect of object oriented refactorings on testability, error proneness and other maintainability attributes, in: Proceedings of the 1st Workshop on Testing Object-Oriented Systems, ACM, 2010, pp. 1--8.

\bibitem{shatnawi2011empirical}
R.~Shatnawi, W.~Li, An empirical assessment of refactoring impact on software quality using a hierarchical quality model, International Journal of Software Engineering and Its Applications 5~(4) (2011) 127--149.

\bibitem{fontana2011impact}
F.~A. Fontana, S.~Spinelli, Impact of refactoring on quality code evaluation, in: Proceedings of the 4th Workshop on Refactoring Tools, 2011, pp. 37--40.

\bibitem{alshayeb2011impact}
M.~Alshayeb, The impact of refactoring to patterns on software quality attributes, Arabian Journal for Science and Engineering 36~(7) (2011) 1241--1251.

\bibitem{lerthathairat2011approach}
P.~Lerthathairat, N.~Prompoon, An approach for source code classification using software metrics and fuzzy logic to improve code quality with refactoring techniques, in: International Conference on Software Engineering and Computer Systems, Springer, 2011, pp. 478--492.

\bibitem{o2012experimental}
M.~{\'O}~Cinn{\'e}ide, L.~Tratt, M.~Harman, S.~Counsell, I.~Hemati~Moghadam, Experimental assessment of software metrics using automated refactoring, in: Proceedings of the ACM-IEEE international symposium on Empirical software engineering and measurement, 2012, pp. 49--58.

\bibitem{ibrahim2012identification}
S.~M. Ibrahim, S.~A. Salem, M.~A. Ismail, M.~Eladawy, Identification of nominated classes for software refactoring using object-oriented cohesion metrics, International Journal of Computer Science Issues (IJCSI) 9~(2) (2012) 68.

\bibitem{singh2011effectiveness}
S.~Singh, K.~S. Kahlon, Effectiveness of encapsulation and object-oriented metrics to refactor code and identify error prone classes using bad smells, ACM SIGSOFT Software Engineering Notes 36~(5) (2011) 1--10.

\bibitem{singh2012effectiveness}
S.~Singh, K.~S. Kahlon, Effectiveness of refactoring metrics model to identify smelly and error prone classes in open source software, ACM SIGSOFT Software Engineering Notes 37~(2) (2012) 1--11.

\bibitem{murgia2012refactoring}
A.~Murgia, R.~Tonelli, M.~Marchesi, G.~Concas, S.~Counsell, J.~McFall, S.~Swift, Refactoring and its relationship with fan-in and fan-out: An empirical study, in: 2012 16th European Conference on Software Maintenance and Reengineering, IEEE, 2012, pp. 63--72.

\bibitem{kannangara2013impact}
S.~Kannangara, W.~Wijayanayake, Impact of refactoring on external code quality improvement: An empirical evaluation, in: 2013 International Conference on Advances in ICT for Emerging Regions (ICTer), IEEE, 2013, pp. 60--67.

\bibitem{veerappa2013empirical}
V.~Veerappa, R.~Harrison, An empirical validation of coupling metrics using automated refactoring, in: 2013 ACM/IEEE International Symposium on Empirical Software Engineering and Measurement, IEEE, 2013, pp. 271--274.

\bibitem{napoli2013using}
C.~Napoli, G.~Pappalardo, E.~Tramontana, Using modularity metrics to assist move method refactoring of large systems, in: 2013 Seventh International Conference on Complex, Intelligent, and Software Intensive Systems, IEEE, 2013, pp. 529--534.

\bibitem{bavota2013empirical}
G.~Bavota, B.~Dit, R.~Oliveto, M.~Di~Penta, D.~Poshyvanyk, A.~De~Lucia, An empirical study on the developers' perception of software coupling, in: Proceedings of the 2013 International Conference on Software Engineering, IEEE Press, 2013, pp. 692--701.

\bibitem{kumari2014effect}
N.~Kumari, A.~Saha, Effect of refactoring on software quality, in: InFourth International Conference on Advances in Computing and Information Technology (ACITY 2014). Delhi, India, 2014.

\bibitem{szoke2014bulk}
G.~Sz{\'o}ke, G.~Antal, C.~Nagy, R.~Ferenc, T.~Gyim{\'o}thy, Bulk fixing coding issues and its effects on software quality: Is it worth refactoring?, in: 2014 IEEE 14th International Working Conference on Source Code Analysis and Manipulation, IEEE, 2014, pp. 95--104.

\bibitem{chaparro2014impact}
O.~Chaparro, G.~Bavota, A.~Marcus, M.~Di~Penta, On the impact of refactoring operations on code quality metrics, in: 2014 IEEE International Conference on Software Maintenance and Evolution, IEEE, 2014, pp. 456--460.

\bibitem{bavota2015experimental}
G.~Bavota, A.~De~Lucia, M.~Di~Penta, R.~Oliveto, F.~Palomba, An experimental investigation on the innate relationship between quality and refactoring, Journal of Systems and Software 107 (2015) 1--14.

\bibitem{kannangara2015empirical}
S.~Kannangara, W.~Wijayanayake, An empirical evaluation of impact of refactoring on internal and external measures of code quality, arXiv preprint arXiv:1502.03526.

\bibitem{gatrell2015effect}
M.~Gatrell, S.~Counsell, The effect of refactoring on change and fault-proneness in commercial c\# software, Science of Computer Programming 102 (2015) 44--56.

\bibitem{cedrim2016does}
D.~Cedrim, L.~Sousa, A.~Garcia, R.~Gheyi, Does refactoring improve software structural quality? a longitudinal study of 25 projects, in: Proceedings of the 30th Brazilian Symposium on Software Engineering, ACM, 2016, pp. 73--82.

\bibitem{malhotra2016empirical}
R.~Malhotra, A.~Chug, An empirical study to assess the effects of refactoring on software maintainability, in: 2016 International Conference on Advances in Computing, Communications and Informatics (ICACCI), IEEE, 2016, pp. 110--117.

\bibitem{mkaouer2016use}
M.~W. Mkaouer, M.~Kessentini, S.~Bechikh, M.~{\'O}. Cinn{\'e}ide, K.~Deb, On the use of many quality attributes for software refactoring: a many-objective search-based software engineering approach, Empirical Software Engineering 21~(6) (2016) 2503--2545.

\bibitem{kaur2017improving}
G.~Kaur, B.~Singh, Improving the quality of software by refactoring, in: 2017 International Conference on Intelligent Computing and Control Systems (ICICCS), IEEE, 2017, pp. 185--191.

\bibitem{chavez2017does}
A.~Ch{\'a}vez, I.~Ferreira, E.~Fernandes, D.~Cedrim, A.~Garcia, How does refactoring affect internal quality attributes?: A multi-project study, in: Proceedings of the 31st Brazilian Symposium on Software Engineering, ACM, 2017, pp. 74--83.

\bibitem{szHoke2017empirical}
G.~Sz{\H{o}}ke, G.~Antal, C.~Nagy, R.~Ferenc, T.~Gyim{\'o}thy, Empirical study on refactoring large-scale industrial systems and its effects on maintainability, Journal of Systems and Software 129 (2017) 107--126.

\bibitem{bashir2017methodology}
R.~S. Bashir, S.~P. Lee, C.~C. Yung, K.~A. Alam, R.~W. Ahmad, A methodology for impact evaluation of refactoring on external quality attributes of a software design, in: 2017 International Conference on Frontiers of Information Technology (FIT), IEEE, 2017, pp. 183--188.

\bibitem{mumtaz2018empirical}
H.~Mumtaz, M.~Alshayeb, S.~Mahmood, M.~Niazi, An empirical study to improve software security through the application of code refactoring, Information and Software Technology 96 (2018) 112--125.

\bibitem{alizadeh2018reducing}
V.~Alizadeh, M.~Kessentini, Reducing interactive refactoring effort via clustering-based multi-objective search, in: 2018 33rd IEEE/ACM International Conference on Automated Software Engineering (ASE), IEEE, 2018, pp. 464--474.

\bibitem{alizadeh2019refbot}
V.~Alizadeh, M.~A. Ouali, M.~Kessentini, M.~Chater, Refbot: intelligent software refactoring bot, in: 2019 34th IEEE/ACM International Conference on Automated Software Engineering (ASE), IEEE, 2019, pp. 823--834.

\bibitem{techapalokul2019code}
P.~Techapalokul, E.~Tilevich, Code quality improvement for all: Automated refactoring for scratch, in: 2019 IEEE Symposium on Visual Languages and Human-Centric Computing (VL/HCC), IEEE, 2019, pp. 117--125.

\bibitem{counsell2019relationship}
S.~Counsell, M.~Arzoky, G.~Destefanis, D.~Taibi, On the relationship between coupling and refactoring: An empirical viewpoint, in: 2019 ACM/IEEE International Symposium on Empirical Software Engineering and Measurement (ESEM), IEEE, 2019, pp. 1--6.

\bibitem{fakhoury2019improving}
S.~Fakhoury, D.~Roy, A.~Hassan, V.~Arnaoudova, Improving source code readability: theory and practice, in: 2019 IEEE/ACM 27th International Conference on Program Comprehension (ICPC), IEEE, 2019, pp. 2--12.

\bibitem{rebai2019interactive}
S.~Rebai, O.~B. Sghaier, V.~Alizadeh, M.~Kessentini, M.~Chater, Interactive refactoring documentation bot, in: 2019 19th International Working Conference on Source Code Analysis and Manipulation (SCAM), IEEE, 2019, pp. 152--162.

\bibitem{alizadeh2019less}
V.~Alizadeh, H.~Fehri, M.~Kessentini, Less is more: From multi-objective to mono-objective refactoring via developer's knowledge extraction, in: 2019 19th International Working Conference on Source Code Analysis and Manipulation (SCAM), IEEE, 2019, pp. 181--192.

\bibitem{alizadeh2018interactive}
V.~Alizadeh, M.~Kessentini, M.~W. Mkaouer, M.~Ocinneide, A.~Ouni, Y.~Cai, An interactive and dynamic search-based approach to software refactoring recommendations, IEEE Transactions on Software Engineering 46~(9) (2018) 932--961.

\bibitem{rebai2020enabling}
S.~Rebai, V.~Alizadeh, M.~Kessentini, H.~Fehri, R.~Kazman, Enabling decision and objective space exploration for interactive multi-objective refactoring, IEEE Transactions on Software Engineering 48~(5) (2020) 1560--1578.

\bibitem{fernandes2020refactoring}
E.~Fernandes, A.~Ch{\'a}vez, A.~Garcia, I.~Ferreira, D.~Cedrim, L.~Sousa, W.~Oizumi, Refactoring effect on internal quality attributes: What haven’t they told you yet?, Information and Software Technology 126 (2020) 106347.

\bibitem{alomar2020developers}
E.~A. AlOmar, P.~T. Rodriguez, J.~Bowman, T.~Wang, B.~Adepoju, K.~Lopez, C.~Newman, A.~Ouni, M.~W. Mkaouer, How do developers refactor code to improve code reusability?, in: International Conference on Software and Software Reuse, Springer, 2020, pp. 261--276.

\bibitem{bibiano2020does}
A.~C. Bibiano, V.~Soares, D.~Coutinho, E.~Fernandes, J.~L. Correia, K.~Santos, A.~Oliveira, A.~Garcia, R.~Gheyi, B.~Fonseca, et~al., How does incomplete composite refactoring affect internal quality attributes?, in: Proceedings of the 28th International Conference on Program Comprehension, 2020, pp. 149--159.

\bibitem{abid2020does}
C.~Abid, M.~Kessentini, V.~Alizadeh, M.~Dhaouadi, R.~Kazman, How does refactoring impact security when improving quality? a security-aware refactoring approach, IEEE Transactions on Software Engineering 48~(3) (2020) 864--878.

\bibitem{abid2021prioritizing}
C.~Abid, V.~Alizadeh, M.~Kessentini, M.~Dhaouadi, R.~Kazman, Prioritizing refactorings for security-critical code, Automated Software Engineering 28~(2) (2021) 1--28.

\bibitem{riansyah2020empirical}
M.~Riansyah, P.~Mursanto, Empirical evaluation of the impact of refactoring on internal quality attributes, in: 2020 International Conference on Advanced Computer Science and Information Systems (ICACSIS), IEEE, 2020, pp. 463--470.

\bibitem{alazzam2020impact}
I.~Alazzam, B.~Abuata, G.~Mhediat, Impact of refactoring on oo metrics: A study on the extract class, extract superclass, encapsulate field and pull up method, International Journal of Machine Learning and Computing 10~(1).

\bibitem{hamdi2021empirical}
O.~Hamdi, A.~Ouni, E.~A. AlOmar, M.~{\'O}. Cinn{\'e}ide, M.~W. Mkaouer, An empirical study on the impact of refactoring on quality metrics in android applications, in: 2021 IEEE/ACM 8th International Conference on Mobile Software Engineering and Systems (MobileSoft), IEEE, 2021, pp. 28--39.

\bibitem{alomar2022refactoring}
E.~A. AlOmar, T.~Wang, V.~Raut, M.~W. Mkaouer, C.~Newman, A.~Ouni, Refactoring for reuse: an empirical study, Innovations in Systems and Software Engineering (2022) 1--31.

\bibitem{sellittotoward}
G.~Sellitto, E.~Iannone, Z.~Codabux, V.~Lenarduzzi, A.~De~Lucia, F.~Palomba, F.~Ferrucci, Toward understanding the impact of refactoring on program comprehension, in: 2022 IEEE international conference on software analysis, evolution and reengineering (SANER), IEEE, 2022, pp. 731--742.

\bibitem{ouni2023impact}
A.~Ouni, E.~A. AlOmar, O.~Hamdi, M.~{\'O}. Cinn{\'e}ide, M.~W. Mkaouer, M.~A. Saied, On the impact of single and co-occurrent refactorings on quality attributes in android applications, Journal of Systems and Software (2023) 111817.

\bibitem{mkaouer2017robust}
M.~W. Mkaouer, M.~Kessentini, M.~{\'O}. Cinn{\'e}ide, S.~Hayashi, K.~Deb, A robust multi-objective approach to balance severity and importance of refactoring opportunities, Empirical Software Engineering 22~(2) (2017) 894--927.

\bibitem{fioravanti2001reengineering}
F.~Fioravanti, G.~Migliarese, P.~Nesi, Reengineering analysis of object-oriented systems via duplication analysis, in: Proceedings of the 23rd International Conference on Software Engineering. ICSE 2001, IEEE, 2001, pp. 577--586.

\bibitem{antoniol2002analyzing}
G.~Antoniol, U.~Villano, E.~Merlo, M.~Di~Penta, Analyzing cloning evolution in the linux kernel, Information and Software Technology 44~(13) (2002) 755--765.

\bibitem{rieger2004insights}
M.~Rieger, S.~Ducasse, M.~Lanza, Insights into system-wide code duplication, in: 11th Working Conference on Reverse Engineering, IEEE, 2004, pp. 100--109.

\bibitem{pantiuchina2020developers}
J.~Pantiuchina, F.~Zampetti, S.~Scalabrino, V.~Piantadosi, R.~Oliveto, G.~Bavota, M.~D. Penta, Why developers refactor source code: A mining-based study, ACM Transactions on Software Engineering and Methodology (TOSEM) 29~(4) (2020) 1--30.

\bibitem{silva2017refdiff}
D.~Silva, M.~T. Valente, Refdiff: detecting refactorings in version histories, in: Proceedings of the 14th International Conference on Mining Software Repositories, IEEE Press, 2017, pp. 269--279.

\bibitem{tsantalis2018accurate}
N.~Tsantalis, M.~Mansouri, L.~M. Eshkevari, D.~Mazinanian, D.~Dig, Accurate and efficient refactoring detection in commit history, in: Proceedings of the 40th International Conference on Software Engineering, ACM, 2018, pp. 483--494.

\bibitem{xiao2024empirical}
L.~Xiao, G.~Zhao, X.~Wang, K.~Li, E.~Lim, C.~Wei, T.~Yu, X.~Wang, An empirical study on the usage of mocking frameworks in apache software foundation, Empirical Software Engineering 29~(2) (2024) 39.

\bibitem{mockus2002two}
A.~Mockus, R.~T. Fielding, J.~D. Herbsleb, Two case studies of open source software development: Apache and mozilla, ACM Transactions on Software Engineering and Methodology (TOSEM) 11~(3) (2002) 309--346.

\bibitem{mockus2000case}
A.~Mockus, R.~T. Fielding, J.~Herbsleb, A case study of open source software development: the apache server, in: Proceedings of the 22nd international conference on Software engineering, 2000, pp. 263--272.

\bibitem{crowston2006assessing}
K.~Crowston, J.~Howison, Assessing the health of open source communities, Computer 39~(5) (2006) 89--91.

\bibitem{rigby2008open}
P.~C. Rigby, D.~M. German, M.-A. Storey, Open source software peer review practices: a case study of the apache server, in: Proceedings of the 30th international conference on Software engineering, 2008, pp. 541--550.

\bibitem{duenas2007apache}
J.~C. Duenas, F.~Cuadrado, M.~Santill{\'a}n, J.~L. Ruiz, et~al., Apache and eclipse: Comparing open source project incubators, IEEE software 24~(6) (2007) 90--98.

\bibitem{weiss2006evolution}
M.~Weiss, G.~Moroiu, P.~Zhao, Evolution of open source communities, in: Open Source Systems: IFIP Working Group 2.13 Foundation on Open Source Software, June 8--10, 2006, Como, Italy 2, Springer, 2006, pp. 21--32.

\bibitem{severance2012apache}
C.~Severance, The apache software foundation: Brian behlendorf, Computer 45~(10) (2012) 8--9.

\bibitem{chidamber1994metrics}
S.~R. Chidamber, C.~F. Kemerer, A metrics suite for object oriented design, IEEE Transactions on software engineering 20~(6) (1994) 476--493.

\bibitem{lorenz1994object}
M.~Lorenz, J.~Kidd, Object-oriented software metrics, Vol. 131, Prentice Hall Englewood Cliffs, 1994.

\bibitem{mccabe1976complexity}
T.~J. McCabe, A complexity measure, IEEE Transactions on software Engineering~(4) (1976) 308--320.

\bibitem{henry1981software}
S.~Henry, D.~Kafura, Software structure metrics based on information flow, IEEE transactions on Software Engineering~(5) (1981) 510--518.

\bibitem{nejmeh1988npath}
B.~A. Nejmeh, Npath: a measure of execution path complexity and its applications, Communications of the ACM 31~(2) (1988) 188--200.

\bibitem{Destefanis:2014:SMA:2813544.2813555}
G.~Destefanis, S.~Counsell, G.~Concas, R.~Tonelli, \href{http://dl.acm.org/citation.cfm?id=2813544.2813555}{Agile processes in software engineering and extreme programming} (2014) 157--170.
\newline\urlprefix\url{http://dl.acm.org/citation.cfm?id=2813544.2813555}

\bibitem{alomar2019can}
E.~A. AlOmar, M.~W. Mkaouer, A.~Ouni, Can refactoring be self-affirmed? an exploratory study on how developers document their refactoring activities in commit messages, in: Proceedings of the 3nd International Workshop on Refactoring-accepted. IEEE, 2019.

\bibitem{alomar2021we}
E.~A. AlOmar, A.~Peruma, M.~W. Mkaouer, C.~Newman, A.~Ouni, M.~Kessentini, How we refactor and how we document it? on the use of supervised machine learning algorithms to classify refactoring documentation, Expert Systems with Applications 167 (2021) 114176.

\bibitem{islam2018characteristics}
M.~R. Islam, M.~F. Zibran, On the characteristics of buggy code clones: A code quality perspective, in: 2018 IEEE 12th International Workshop on Software Clones (IWSC), IEEE, 2018, pp. 23--29.

\bibitem{singh2012evaluation}
V.~Singh, V.~Bhattacherjee, Evaluation and application of package level metrics in assessing software quality, Vol.~58, Foundation of Computer Science, 2012.

\bibitem{wilcoxon1945individual}
F.~Wilcoxon, Individual comparisons by ranking methods, Biometrics bulletin 1~(6) (1945) 80--83.

\bibitem{trove.nla.gov.au/work/16432558}
R.~J. Grissom, J.~J. Kim, Effect sizes for research : a broad practical approach, Mahwah, N.J. ; London : Lawrence Erlbaum Associates, 2005, formerly CIP.

\bibitem{henderson1995object}
B.~Henderson-Sellers, Object-oriented metrics: measures of complexity, Prentice-Hall, Inc., 1995.

\bibitem{higo2004aries}
Y.~Higo, T.~Kamiya, S.~Kusumoto, K.~Inoue, K.~Words, Aries: Refactoring support environment based on code clone analysis., in: IASTED Conf. on Software Engineering and Applications, 2004, pp. 222--229.

\bibitem{higo2005aries}
Y.~Higo, T.~Kamiya, S.~Kusumoto, K.~Inoue, Aries: refactoring support tool for code clone, ACM SIGSOFT Software Engineering Notes 30~(4) (2005) 1--4.

\bibitem{higo2008metric}
Y.~Higo, S.~Kusumoto, K.~Inoue, A metric-based approach to identifying refactoring opportunities for merging code clones in a {Java} software system, Journal of Software Maintenance and Evolution: Research and Practice 20~(6) (2008) 435--461.

\bibitem{tairas2012increasing}
R.~Tairas, J.~Gray, Increasing clone maintenance support by unifying clone detection and refactoring activities, Information and Software Technology 54~(12) (2012) 1297--1307.

\bibitem{bian2013spape}
Y.~Bian, G.~Koru, X.~Su, P.~Ma, Spape: A semantic-preserving amorphous procedure extraction method for near-miss clones, Journal of Systems and Software 86~(8) (2013) 2077--2093.

\bibitem{yue2018automatic}
R.~Yue, Z.~Gao, N.~Meng, Y.~Xiong, X.~Wang, J.~D. Morgenthaler, Automatic clone recommendation for refactoring based on the present and the past, in: 2018 IEEE International Conference on Software Maintenance and Evolution (ICSME), IEEE, 2018, pp. 115--126.

\bibitem{yoshida2019proactive}
N.~Yoshida, S.~Numata, E.~Choiz, K.~Inoue, {Proactive clone recommendation system for Extract Method refactoring}, in: 2019 IEEE/ACM 3rd International Workshop on Refactoring (IWoR), IEEE, 2019, pp. 67--70.

\bibitem{arcelli2015duplicated}
F.~Arcelli~Fontana, M.~Zanoni, F.~Zanoni, A duplicated code refactoring advisor, in: Agile Processes in Software Engineering and Extreme Programming: 16th International Conference, XP 2015, Helsinki, Finland, May 25-29, 2015, Proceedings 16, Springer, 2015, pp. 3--14.

\bibitem{alomar2022anticopypaster}
E.~A. AlOmar, A.~Ivanov, Z.~Kurbatova, Y.~Golubev, M.~W. Mkaouer, A.~Ouni, T.~Bryksin, L.~Nguyen, A.~Kini, A.~Thakur, Anticopypaster: Extracting code duplicates as soon as they are introduced in the ide, in: 37th IEEE/ACM International Conference on Automated Software Engineering, 2022, pp. 1--4.

\bibitem{alomar2023just}
E.~A. AlOmar, A.~Ivanov, Z.~Kurbatova, Y.~Golubev, M.~W. Mkaouer, A.~Ouni, T.~Bryksin, L.~Nguyen, A.~Kini, A.~Thakur, Just-in-time code duplicates extraction, Information and Software Technology (2023) 107169.

\bibitem{alomar2024behind}
E.~A. AlOmar, M.~W. Mkaouer, A.~Ouni, Behind the intent of extract method refactoring: A systematic literature review, IEEE Transactions on Software Engineering.

\bibitem{griswold1993automated}
W.~G. Griswold, D.~Notkin, Automated assistance for program restructuring, ACM Transactions on Software Engineering and Methodology (TOSEM) 2~(3) (1993) 228--269.

\bibitem{mondal2020survey}
M.~Mondal, C.~K. Roy, K.~A. Schneider, A survey on clone refactoring and tracking, Journal of Systems and Software 159 (2020) 110429.

\bibitem{murphy2012we}
E.~Murphy-Hill, C.~Parnin, A.~P. Black, How we refactor, and how we know it, IEEE Transactions on Software Engineering 38~(1) (2012) 5--18.

\bibitem{nguyen2013filtering}
H.~A. Nguyen, A.~T. Nguyen, T.~N. Nguyen, Filtering noise in mixed-purpose fixing commits to improve defect prediction and localization, in: 2013 IEEE 24th international symposium on software reliability engineering (ISSRE), IEEE, 2013, pp. 138--147.

\bibitem{baqais2020automatic}
A.~A.~B. Baqais, M.~Alshayeb, Automatic software refactoring: a systematic literature review, Software Quality Journal 28~(2) (2020) 459--502.

\bibitem{hall2012supervised}
M.~Hall, N.~Walkinshaw, P.~McMinn, Supervised software modularisation, in: 2012 28th IEEE International Conference on Software Maintenance (ICSM), IEEE, 2012, pp. 472--481.

\bibitem{bavota2012putting}
G.~Bavota, F.~Carnevale, A.~De~Lucia, M.~Di~Penta, R.~Oliveto, Putting the developer in-the-loop: An interactive ga for software re-modularization, in: Search Based Software Engineering: 4th International Symposium, SSBSE 2012, Riva del Garda, Italy, September 28-30, 2012. Proceedings 4, Springer, 2012, pp. 75--89.

\bibitem{herzig2016impact}
K.~Herzig, S.~Just, A.~Zeller, The impact of tangled code changes on defect prediction models, Empirical Software Engineering 21~(2) (2016) 303--336.

\bibitem{kirinuki2014hey}
H.~Kirinuki, Y.~Higo, K.~Hotta, S.~Kusumoto, Hey! are you committing tangled changes?, in: Proceedings of the 22nd International Conference on Program Comprehension, 2014, pp. 262--265.

\end{thebibliography}

\end{document}